\documentclass[fleqn,usenatbib]{mnras}

\usepackage{newtxtext,newtxmath}
\usepackage{ae,aecompl}

\usepackage[T1]{fontenc}

\DeclareRobustCommand{\VAN}[3]{#2}
\let\VANthebibliography\thebibliography
\def\thebibliography{\DeclareRobustCommand{\VAN}[3]{##3}\VANthebibliography}

\usepackage{graphicx}	
\usepackage{amsmath}	

\newcommand{\kms}{{\rm\,km\,s^{-1}}}

\title[Orbits of globular clusters with dynamical friction]{Orbits of 
globular clusters computed with dynamical friction in the Galactic
anisotropic velocity dispersion field}

\author[E. Moreno et al.]{
Edmundo Moreno,$^{1}$\thanks{E-mail: edmundo@astro.unam.mx}
Jos\'e G. Fern\'andez-Trincado,$^{2}$
Angeles P\'erez-Villegas,$^{3}$\thanks{E-mail: mperez@astro.unam.mx}
Leonardo Chaves-Velasquez$^{4,5,6}$
\newauthor
\,William J. Schuster,$^{3}$
\\
$^{1}$Instituto de Astronom\'ia, Universidad Nacional Aut\'onoma de
M\'exico, Apdo. Postal 70-264, Ciudad Universitaria CDMX 04510,
M\'exico\\
$^{2}$Instituto de Astronom\'ia, Universidad Cat\'olica del Norte, Av.
Angamos 0610, Antofagasta, Chile \\
$^{3}$Instituto de Astronom\'ia, Universidad Nacional Aut\'onoma de
M\'exico, Apdo. Postal 106, 22800 Ensenada, B.C., M\'exico\\
$^{4}$Instituto de Radioastronom\'ia y Astrof\'isica, Universidad Nacional Aut\'onoma de M\'exico, Apdo. Postal 3-72, Morelia Michoac\'an, 58089, M\'exico\\
$^{5}$Astronomical Observatory, Universidad de Nari\~no, Sede VIIS, Avenida Panamericana, Pasto, 520017 Nari\~no, Colombia\\
$^{6}$Departamento de F\'isica de la Universidad de Nari\~no, Torobajo Calle 18 Carrera 50, Pasto, 520017 Nari\~no, Colombia
}

\date{Accepted XXX. Received YYY; in original form ZZZ}

\pubyear{2021}

\begin{document}
\label{firstpage}
\pagerange{\pageref{firstpage}--\pageref{lastpage}}
\maketitle

\begin{abstract}

We present a preliminary analysis of the effect of dynamical friction
on the orbits of part of the globular clusters in our Galaxy.
Our study considers an anisotropic velocity dispersion field
approximated using the results of studies in the literature.
An axisymmetric Galactic model with
mass components consisting of a disc, a bulge, and a dark halo is
employed in the computations. We provide a method to
compute the dynamical friction acceleration in ellipsoidal, oblate,
and prolate velocity distribution functions with similar density in
velocity space. Orbital properties, such as mean time-variations of
perigalactic and apogalactic distances, energy, and z-component of 
angular momentum, are obtained for globular clusters
lying in the Galactic region $R \lesssim$ 10 kpc, $|z| \lesssim$ 5 kpc,
with $R,z$ cylindrical coordinates. These include clusters in 
prograde and retrograde orbital motion. Several clusters are strongly
affected by dynamical friction, in particular Liller 1,
Terzan 4, Terzan 5, NGC 6440, and NGC 6553, which lie in the
Galactic inner region. We comment on the more relevant
implications of our results on the
dynamics of Galactic globular clusters, such as their possible
misclassification between the categories \lq halo',
\lq bulge', and \lq thick disc', the resulting biasing of
 globular-cluster
samples, the possible incorrect association of the globulars with
their parent dwarf galaxies for accretion events, and the possible
formation of \lq nuclear star clusters'.
\end{abstract}




\section{Introduction}

The acceleration due to dynamical friction on a massive object moving
through an infinite, homogeneous stellar medium was obtained
by \citet{1943ApJ....97..255C}, assuming an isotropic velocity
dispersion of background stars.
\citet{1975ApJ...196..407T} employed this result to study the possible
formation of nuclei of galaxies when globular clusters spiral into
their galactic centre, as they lose orbital energy due to dynamical
friction. In our Galaxy, approximating the outer regions with an
isotropic velocity dispersion, \citet{1976ApJ...203...72T} computed
the orbit of the Large Magellanic Cloud under the dynamical friction
effect. A detailed study of this effect on the motion of an object
with mass 7 $\times 10^5$ M$_\odot$, moving close to the disc of a
Galactic potential model, was made by \citet{1979A&A....71..245K}.  
He used the \lq representative' Galactic model of
\citet{1973Ap&SS..22..393I}, taking also constant values of isotropic
velocity dispersions in each of the four oblate spheroids representing
the disc component in the model.

 Employing the general case of an anisotropic velocity
dispersion field, several studies analyse the evolution of orbits due
to dynamical friction in a given potential. A spherical potential has
been considered by \citet{1987IAUS..127..475C} and
\citet{2000ApJ...532..294T}. An oblate potential is employed by 
\citet{1977SvA....21...12S} and \citet{1988ApJ...331...71S}.  
The motion in triaxial potentials has been studied by
\citet{1991ApJ...375..544S} and \citet{1992MNRAS.254..466P}. 
In the present analysis we develop a method to compute the
dynamical friction acceleration produced by velocity distribution
functions whose constant density surfaces in velocity space are
ellipsodal, oblate, or prolate similar surfaces, i.e. they have
constant semiaxes ratios. We apply this theory in the disc, bulge,
and dark halo of our Galaxy to compute orbits of globular clusters.
Recent data obtained by the $Gaia$ mission have increased our
understanding of the Galactic kinematics, including in particular
the kinematics of the system of globular clusters \citep{2018A&A...616A..11G,2018A&A...616A..12G},
which have been well complemented with large spectroscopic surveys such 
as APOGEE-2/SDSS-IV \citep{2017AJ....154...94M}.
From the resulting analysis of these data some new accretion events
in our Galaxy have been inferred: Gaia-Sausage \citep{2018MNRAS.478..611B}, Gaia-Enceladus \citep{2018Natur.563...85H}, the possible shards of 
$\omega$ Centauri \citep{2018MNRAS.478.5449M}, a blob in the nearby
stellar halo and a stellar population on highly eccentric 
orbits \citep{2018ApJ...860L..11K,2019MNRAS.482.3426M},  
Sequoia \citep{2019MNRAS.488.1235M}, 
Kraken or Koala \citep{2019MNRAS.486.3180K,2020MNRAS.498.2472K,2020MNRAS.493..847F},
and a number of chemically
anomalous structures associated with the disruption of globular
clusters \citep[see, e.g.,][]{2016ApJ...833..132F,2017ApJ...846L...2F,2019MNRAS.488.2864F,2019ApJ...886L...8F,2020MNRAS.495.4113F,2020A&A...643L...4F,2020A&A...644A..83F,2021ApJ...918L..37F,2021A&A...647A..64F,2021A&A...648A..70F}.
These events significantly increase the number of already known events
of this type: 
Sagittarius \citep{1994Natur.370..194I}, Helmi Stream \citep{1999Natur.402...53H},
Canis Major dwarf galaxy \citep{2004MNRAS.348...12M},
a substructure on the Galactic disc \citep{2006MNRAS.365.1309H}.
All this information
has led to the conclusion that some globular clusters have been
accreted in these events, and others have been formed in situ
\citep{2003AJ....125..188B,2010MNRAS.404.1203F,2018ApJ...863L..28M,
2018Natur.563...85H,2019MNRAS.488.1235M,2019A&A...630L...4M,
2019A&A...625A...5K,2020MNRAS.493..847F}.

In the process to associate some clusters to a given accretion
event, it has been assumed that the orbital energy and $z$-component of
the angular momentum are constants, or vary slightly, in time.
Under the action of dynamical friction and employing an axisymmetric
Galactic potential, these orbital properties are not strictly constant.
Thus, the associations cluster-accretion event could be uncertain.

In our analysis of dynamical friction, and in particular for 
globular clusters whose orbits lie in an inner Galactic region, we
quantify the variation of orbital energy and $z$ angular momentum. 
Our results show that according to the \textit{current} values of these
variations, the assumption of constant orbital properties 
is approximately appropriate for the majority of studied clusters,
except in some cases which show strong variations. However, a
further analysis is needed computing the orbits with a backward
integration in time, considering an increase of mass of the clusters
\citep{2019MNRAS.482.5138B}. As the
acceleration due to dynamical friction depends directly on this mass,
the expected greater mass in the past in combination with the
decrease of the background density of Galactic regions where the
cluster moved, could result in variations of energy and angular
momentum similar to the current ones, but some particular clusters
deserve a detailed analysis.

This work is organised as follows: In Section 2 and the
 Appendices, we describe our
method to compute the acceleration due to dynamical friction.
Section 3 gives the approximated anisotropic velocity dispersion
field employed in the computations, using results from the
literature. In Section 4, we compute the orbits
of globular clusters that lie within a restricted Galactic region.
Implications of our results for the system of Galactic globular clusters
are commented in Section 5. Our conclusions are given in Section 6.

\section{Dynamical friction acceleration}

\label{acel}

The dynamical-friction acceleration of a body of mass $M$ moving
with instantaneous velocity {\boldmath $v$}$_{\rm M}$ with respect
to the local velocity centroid, at position {\boldmath $r$},
 of a system of perturbing particles of
mass $m_{\rm p} \ll M$ is \citep{BT08,SP87}

\begin{equation} \left(\frac{{\rm d}\mbox{\boldmath $v$}_M}{{\rm d} t} \right)_{\rm df} = 4{\pi}G^2Mm_{\rm p}\ln\Lambda {\nabla}_{\tiny \mbox{\boldmath $v$}_{\tiny \rm M}}h(\mbox{\boldmath $r$},\mbox{\boldmath $v$}_{\rm M}),
\label{acfd}
\end{equation}

\noindent with $\ln\Lambda$ the Coulomb logarithm commented below, and
 $h$({\boldmath $r$},{\boldmath $v$}$_{\rm M}$) the Rosenbluth
 potential \citep{1957PhRv..107....1R} given by

\begin{equation} h(\mbox{\boldmath $r$},\mbox{\boldmath $v$}_{\rm M}) =
\int \frac{f(\mbox{\boldmath $r$},\mbox{\boldmath $v$})}{|\mbox{\boldmath $v$}_{\rm M}-
\mbox{\boldmath $v$}|}{d}^3\mbox{\boldmath $v$}.
\label{fh}
\end{equation}

\noindent $f$({\boldmath $r$},{\boldmath $v$}) is the
local distribution function (DF) of particles $m_{\rm p}$, with their
 velocity
{\boldmath $v$} given with respect to their local centroid.
The subscript {\boldmath $v$}$_{\rm M}$ in the gradient operator
${\nabla}$ in Eq.~\ref{acfd} denotes derivation in velocity space.

In the acceleration given in Eq.~\ref{acfd}, we consider the
contribution of particles $m_{\rm p}$ in the three Galactic components:
bulge, disc, and dark halo. In each component we assume a local
DF of the type

\begin{equation} f(\mbox{\boldmath $r$},\mbox{\boldmath $v$}) =
\frac{n}{(2\pi)^{3/2}\sigma_1\sigma_2\sigma_3}\exp\left[-\frac{1}{2}
\left(\frac{{v}_1^2}{\sigma_1^2}+
\frac{{v}_2^2}{\sigma_2^2}+
\frac{{v}_3^2}{\sigma_3^2}\right)\right],
\label{fd}
\end{equation}

\noindent with $n$, $(\sigma_1,\sigma_2,\sigma_3)$,
$({v}_1,{v}_2,{v}_3)$ respectively the
 corresponding
local number density, velocity dispersions, and velocity components
along the principal axes of the distribution.
 In our analysis, we consider
the general case in which $(\sigma_1,\sigma_2,\sigma_3)$ may be
different, i.e. an anisotropic dispersion field. With this DF and
defining

\begin{equation} f_0(\mbox{\boldmath $r$},\mbox{\boldmath $v$})
=\exp\left[-\frac{1}{2}\left(\frac{{v}_1^2}{\sigma_1^2}+
\frac{{v}_2^2}{\sigma_2^2}+
\frac{{v}_3^2}{\sigma_3^2}\right)\right],
\label{fd0}
\end{equation}

\begin{equation} {\Psi}(\mbox{\boldmath $r$},\mbox{\boldmath $v$}_{\rm M}) =
-\int \frac{f_0(\mbox{\boldmath $r$},\mbox{\boldmath $v$})}{|\mbox{\boldmath $v$}_{\rm M}-
\mbox{\boldmath $v$}|}{d}^3\mbox{\boldmath $v$},
\label{fh0}
\end{equation}

\noindent then Eq.~\ref{acfd} is

\begin{equation} \left(\frac{{\rm d}\mbox{\boldmath $v$}_M}{{\rm d}t} \right)_{\rm df} = - \sqrt{\frac{2}{\pi}}\frac{G^2M\rho_{\rm p}\ln\Lambda}
{\sigma_1\sigma_2\sigma_3} {\nabla}_{\tiny \mbox{\boldmath $v$}_{\tiny \rm M}}{\Psi}(\mbox{\boldmath $r$},\mbox{\boldmath $v$}_{\rm M}),
\label{acfdnu}
\end{equation}

\noindent with $\rho_{\rm p}$ the corresponding local mass density of
particles $m_{\rm p}$.

The density function $f_0$({\boldmath $r$},{\boldmath $v$})
is constant on similar ellipsoidal surfaces in velocity space, and can
 be expressed as
a function of an appropriate variable identifying these surfaces.
For the computation of the force field
$-{\nabla}$$_{\tiny v_{\rm M}}$${\Psi}$({\boldmath $r$},{\boldmath $v$}$_{\rm M}$) in Eq.~\ref{acfdnu}, we find it convenient to
approximate $f_0$ with a series of connected linear segments, which
are linear functions of the employed variable. Each linear segment
represents a shell in velocity space. In the
Appendices~\ref{ap1},\ref{ap2},\ref{ap3} we present a method
to compute in velocity space the force field of a shell of
this type, for ellipsoidal: $\sigma_1$>$\sigma_2$>$\sigma_3$, oblate:
$\sigma_1$=$\sigma_2$>$\sigma_3$, and prolate:
$\sigma_1$>$\sigma_2$=$\sigma_3$ arguments of the function $f_0$.
In these Appendices the point {\boldmath $v$}=$({v}_1,{v}_2,{v}_3)$
represents the velocity {\boldmath $v$}$_{\rm M}$ expressed in
the employed Cartesian base ({\boldmath $e$}$_1$,{\boldmath $e$}$_2$,{\boldmath $e$}$_3$)

\begin{equation}
 \mbox{\boldmath $v$}_{\tiny \rm M}=
\mbox{\boldmath $e$}_1{v}_{\rm M1}+
\mbox{\boldmath $e$}_2{v}_{\rm M2}+
\mbox{\boldmath $e$}_3{v}_{\rm M3}.
\label{vM}
\end{equation}

The total field $-{\nabla}$$_{\tiny v_{\rm M}}$${\Psi}$ at this point
{\boldmath $v$}$_{\rm M}$ is obtained with the sum of the
individual fields of each shell. As stated in parts (b) and (c) of
Appendices~\ref{ap1},\ref{ap2},\ref{ap3}, the force field inside a
shell is zero; thus,
in that procedure we take shells only inside the similar surface
which crosses the point {\boldmath $v$}$_{\rm M}$.

In the isotropic case $\sigma_1$=$\sigma_2$=$\sigma_3$=$\sigma$, and
with $\gamma$=|{\boldmath $v$}$_{\rm M}$|$/(\sqrt{2}\sigma)$ we
have

\begin{equation}
 -{\nabla}_{\tiny \mbox{\boldmath $v$}_{\tiny \rm M}}{\Psi}(\mbox{\boldmath $r$},\mbox{\boldmath $v$}_{\rm M})=
-\mbox{\boldmath $v$}_{\rm M}\left(\frac{\sqrt{\pi}}{\gamma}\right)^3\left({\rm erf}(\gamma)-\frac{2}{\sqrt{\pi}}\gamma{\rm e}^{-{\gamma}^2}\right).
\label{grad}
\end{equation}

In Section~\ref{dispvobs}, we will take an isotropic bulge component, 
and assume that at any point {\boldmath $r$} in the dark halo
component, the 
principal axes of the corresponding DF lie along the directions of
 unitary vectors
{\boldmath $e$}$_r$,{\boldmath $e$}$_{\varphi}$,
{\boldmath $e$}$_{\theta}$ in spherical coordinates. In the disc
component the principal axes will be taken along the directions of
unitary vectors {\boldmath $e$}$_R$,{\boldmath $e$}$_z$,{\boldmath $e$}$_{\varphi}$
in cylindrical coordinates. Thus, in
the dark halo and disc components the corresponding local dispersions
$\sigma_r,\sigma_{\varphi},\sigma_{\theta}$ and
$\sigma_R,\sigma_z,\sigma_{\varphi}$ are arranged as
$\sigma_1,\sigma_2,\sigma_3$ with
$\sigma_1$$\geq$$\sigma_2$$\geq$$\sigma_3$, and set the corresponding
right-handed Cartesian base
({\boldmath $e$}$_1$,{\boldmath $e$}$_2$,{\boldmath $e$}$_3$) to be
employed in the Appendices~\ref{ap1},\ref{ap2},\ref{ap3}.
 In Table~\ref{cases}, we give the
possible cases in this arrangement and the assumed base
({\boldmath $e$}$_1$,{\boldmath $e$}$_2$,{\boldmath $e$}$_3$
). For a given base, the components
${v}_{\rm M1}$,${v}_{\rm M2}$,${v}_{\rm M2}$ in
Eq.~\ref{vM} are obtained assuming that in each Galactic component the
velocity of the local centroid with respect to the Galactic inertial
frame points in the {\boldmath $e$}$_{\varphi}$ direction, this
velocity being <${v}_{\varphi}$>{\boldmath $e$}$_{\varphi}$.
Thus, with principal axes in spherical coordinates, if
${V}_r$,${V}_{\varphi}$,${V}_{\theta}$ are the
components of the velocity {\boldmath $V$} of body $M$ with respect
to the Galactic inertial frame, {\boldmath $v$}$_{\rm M}$
for the given Galactic component is

\begin{equation}
 \mbox{\boldmath $v$}_{\tiny \rm M}=
\mbox{\boldmath $e$}_r{V}_r+
\mbox{\boldmath $e$}_{\varphi}({V}_{\varphi}-<\!{v}_{\varphi}\!>)+
\mbox{\boldmath $e$}_{\theta}{V}_{\theta},
\label{vMesf}
\end{equation}

\noindent and with principal axes in cylindrical coordinates, and
${V}_R$,${V}_z$,${V}_{\varphi}$ the corresponding components of
{\boldmath $V$}, the velocity {\boldmath $v$}$_{\rm M}$ is

\begin{equation}
 \mbox{\boldmath $v$}_{\tiny \rm M}=
\mbox{\boldmath $e$}_R{V}_R+
\mbox{\boldmath $e$}_z{V}_z+
\mbox{\boldmath $e$}_{\varphi}({V}_{\varphi}-<\!{v}_{\varphi}\!>).
\label{vMcil}
\end{equation}

In both situations the components
${v}_{\rm M1}$,${v}_{\rm M2}$,${v}_{\rm M3}$
follow inserting a corresponding base
({\boldmath $e$}$_1$,{\boldmath $e$}$_2$,{\boldmath $e$}$_3$)
in Eq.~\ref{vM} from Table~\ref{cases} and comparing with
Eq.~\ref{vMesf} or Eq.~\ref{vMcil}. The total field
$-{\nabla}$$_{\tiny v_{\rm M}}$${\Psi}$
in Eq.~\ref{acfdnu} is expressed in this base
({\boldmath $e$}$_1$,{\boldmath $e$}$_2$,{\boldmath $e$}$_3$).

With $\Phi$ the total Galactic potential, the instantaneous
acceleration of $M$ with respect to the Galactic inertial frame is

\begin{equation} \frac{{\rm d}\mbox{\boldmath $V$}}{{\rm d}t}=
-\nabla\Phi+\sum_{\rm i}\left(\frac{{\rm d}\mbox{\boldmath $v$}_M}{{\rm d}t} \right)_{\rm df_i},
\label{actot}
\end{equation}

\noindent the second term giving the contributions of the three
Galactic components: bulge, disc, and dark halo.

\begin{table}
 \caption{Cases in the arrangement of local dispersions.}
 \label{cases}
 \begin{tabular}{ccccc}
  \hline
 & \multicolumn{2}{c}{spherical} & \multicolumn{2}{c}{cylindrical} \\
 Case & ($\sigma_1,\sigma_2,\sigma_3$)  & ({\boldmath $e$}$_1$, {\boldmath $e$}$_2$, {\boldmath $e$}$_3$) &
      ($\sigma_1,\sigma_2,\sigma_3$)    & ({\boldmath $e$}$_1$, {\boldmath $e$}$_2$, {\boldmath $e$}$_3$) \\
  \hline
  1 & ($\sigma_r,\sigma_{\theta},\sigma_{\varphi}$) & ({\boldmath $e$}$_r$, {\boldmath $e$}$_{\theta}$, {\boldmath $e$}$_{\varphi}$) & ($\sigma_z,\sigma_R,\sigma_{\varphi}$)  & ({\boldmath $e$}$_z$, {\boldmath $e$}$_R$, {\boldmath $e$}$_{\varphi}$)  \\
  2 & ($\sigma_r,\sigma_{\varphi},\sigma_{\theta}$) & ({\boldmath $e$}$_r$, {\boldmath $e$}$_{\varphi}$, $-${\boldmath $e$}$_{\theta}$) & ($\sigma_z,\sigma_{\varphi},\sigma_R$)  & ({\boldmath $e$}$_z$, {\boldmath $e$}$_{\varphi}$, $-${\boldmath $e$}$_R$)   \\
  3 & ($\sigma_{\theta},\sigma_{\varphi},\sigma_r$) & ({\boldmath $e$}$_{\theta}$, {\boldmath $e$}$_{\varphi}$, {\boldmath $e$}$_r$) & ($\sigma_R,\sigma_{\varphi},\sigma_z$)  & ({\boldmath $e$}$_R$, {\boldmath $e$}$_{\varphi}$, {\boldmath $e$}$_z$)    \\
  4 & ($\sigma_{\theta},\sigma_r,\sigma_{\varphi}$) & ({\boldmath $e$}$_{\theta}$, {\boldmath $e$}$_r$, $-${\boldmath $e$}$_{\varphi}$) & ($\sigma_R,\sigma_z,\sigma_{\varphi}$)  & ({\boldmath $e$}$_R$, {\boldmath $e$}$_z$, $-${\boldmath $e$}$_{\varphi}$)   \\
  5 & ($\sigma_{\varphi},\sigma_r,\sigma_{\theta}$) & ({\boldmath $e$}$_{\varphi}$, {\boldmath $e$}$_r$, {\boldmath $e$}$_{\theta}$) & ($\sigma_{\varphi},\sigma_z,\sigma_R$)  & ({\boldmath $e$}$_{\varphi}$, {\boldmath $e$}$_z$, {\boldmath $e$}$_R$)    \\
  6 & ($\sigma_{\varphi},\sigma_{\theta},\sigma_r$) & ({\boldmath $e$}$_{\varphi}$, {\boldmath $e$}$_{\theta}$, $-${\boldmath $e$}$_r$) & ($\sigma_{\varphi},\sigma_R,\sigma_z$)  & ({\boldmath $e$}$_{\varphi}$, {\boldmath $e$}$_R$, $-${\boldmath $e$}$_z$)   \\
  \hline
 \end{tabular}
\end{table}

In the computation of Galactic orbits of globular clusters, following
\citet{BT08}, at every orbital point the factor
$\ln\Lambda$ in Eqs.~\ref{acfd},~\ref{acfdnu} is approximated as

\begin{equation} \ln\Lambda= \ln \left (\frac{b_{\rm max}}{{\rm max} \left[r_{\rm h},GM/(\sigma_1\sigma_2\sigma_3)^{2/3}\right]} \right ),
\label{faclam}
\end{equation}

\noindent with $r_{\rm h}$ being the half-mass radius of the cluster,
and the maximum impact parameter $b_{\rm max}$ is set equal to the
instantaneous distance from the cluster to the Galactic Centre.

\section{Velocity dispersions and mean rotation velocity of mass
components in our Galaxy}
\label{dispvobs}

To analyse the effect of dynamical friction in our Galaxy, the velocity
dispersions and the mean rotation velocity of the mass components are
approximated with some studies in the literature. These velocity fields
are employed in an axisymmetric Galactic potential model. The considered
mass components are a disc, a bulge, and a dark halo; in the following, 
we summarise some of their properties, along with the Galactic 
potential.
Due to the convenient use in the computations of
analytic expressions relating the velocity dispersions and mean rotation
velocity with a given position in the Galaxy, we will consider results
from the literature, which provide approximately analytic details for
these properties in the disc, bulge, and dark halo. 

In this preliminary analysis of the effect of dynamical
friction, we compute the motion of those globular clusters whose orbits 
lie within the restricted Galactic region defined approximately by
$R \lesssim$ 10 kpc, $|z| \lesssim$ 5 kpc, with $R,z$ cylindrical
coordinates. The consideration of this restricted region is due to the
approximated velocity dispersions and mean rotation velocity employed
in the disc component, whose details are given in the following
section. Thus, for the three mass components, we focus on the velocity
fields within this region. Future analyses will consider a more
extended Galactic region.

\subsection{Disc component}
\label{disco}

The Galactic Disc has been analysed in several studies that include
the thin or/and thick discs, e.g. \citet{
1988A&A...192..117V,1989AJ.....97..139L,
1989ARA&A..27..555G,1999A&A...352..129V,2003A&A...409..523R,
2010ApJ...716....1B,
2010ApJ...712..692C,2010A&A...510L...4S,
2010gama.conf..153V, 2011ApJ...738..187L,
2011ARA&A..49..301V, 2012A&A...547A..70P,
2012A&A...547A..71P, 2014A&A...569A..13R,
2014ApJ...793...51S,2015A&A...583A..91G,
2018A&A...616A..11G, 2020MNRAS.494.6001N,
2020A&A...634A..66L, 2021MNRAS.506.1761S}.
In this work, we consider
the relations of the form $a+b|z|^c$ for the 
velocity dispersions $\sigma_R$,$\sigma_z$,$\sigma_{\varphi}$, and mean velocity
<${v}_{\varphi}$>, given by \citet{2010ApJ...716....1B};
these approximations apply 
at $R_0$, the Galactocentric position of the Sun, and towards the 
north Galactic Pole. In these relations the value of the distance
from the plane, $|z|$, is given in the interval 
<1,5> kpc, and the dispersions and mean velocity result in $\kms$.
In Table~\ref{disco} we list the values of the parameters
$a,b,c$ given by \citet{2010ApJ...716....1B}. In our computations a
negative value of <${v}_{\varphi}$> indicates prograde motion,
i.e. in the sense of Galactic rotation; see Section~\ref{modgal} for
the reference frame taken in the computations. Figures 12 and 13
in \citet{2015A&A...583A..91G} show that there is no great variation
of dispersions and mean velocity in $|z|$ < 1 kpc, thus the relations
 $a+b|z|^c$
of \citet{2010ApJ...716....1B} can be approximately applied towards
$z$=0. More detailed relations for $\sigma_R$,$\sigma_z$ have been
given by \citet{2021MNRAS.506.1761S}, which could be employed in 
a further analysis.

Following the results obtained by \citet{1989AJ.....97..139L},
in cylindrical coordinates $(R,z,\varphi)$ we take
for the velocity dispersion $\sigma_R(R,z)$ the form

\begin{equation} \sigma_R(R,z)=\sigma_0{\rm e}^{-R/2h_R}\sigma_R(R_0,z),
\label{disRd1}
\end{equation}
 
\noindent with $h_R$ being a scale length and $\sigma_R(R_0,z)$ obtained
with Table~\ref{disco}. Evaluating at $R_0$, this
gives

\begin{equation} \sigma_R(R,z)={\rm e}^{-(R-R_0)/2h_R}\sigma_R(R_0,z).
\label{disRd}
\end{equation}

Analogously, 

\begin{equation} \sigma_z(R,z)={\rm e}^{-(R-R_0)/2h_z}\sigma_z(R_0,z),
\label{diszd}
\end{equation}

\begin{equation} \sigma_{\varphi}(R,z)={\rm e}^{-(R-R_0)/2h_{\varphi}}\sigma_{\varphi}(R_0,z).
\label{disfd}
\end{equation}

In our computations we take the values $h_R$=$h_z$=4.37 kpc,
$h_{\varphi}$=3.36 kpc given by \citet{1989AJ.....97..139L}.

To approximate <${v}_{\varphi}$>$(R,z)$, figure 13 given by
\citet{2018A&A...616A..11G} suggests that we can approximate the
rotation velocity at any level
$z$ with a Brandt velocity function \citep{1960ApJ...131..293B}

\begin{equation} V(R)=\frac{3V_{\rm max}\left(\frac{R}{R_{\rm max}}\right)}{1+2\left(\frac{R}{R_{\rm max}}\right)^{3/2}},
\label{brandt}
\end{equation}

\noindent with $R_{\rm max}$ the position where the maximum value
$V_{\rm max}$ is reached. In terms of $R_0$ and the corresponding
velocity $V(R_0)$=<${v}_{\varphi}$>$(R_0,z)$, the mean rotation
velocity is given by

\begin{equation} <\!{v}_{\varphi}\!>(R,z)=
<\!{v}_{\varphi}\!>(R_0,z)\left(\frac{R}{R_0}\right)\frac{1+2\left(\frac{R_0}{R_{\rm max}}\right)^{3/2}}{1+2\left(\frac{R}{R_{\rm max}}\right)^{3/2}}.
\label{vfd}
\end{equation}

Figure 13 in \citet{2018A&A...616A..11G} shows that $R_{\rm max}$
lies approximately in the interval 6--8 kpc; in this work we take
$R_{\rm max}$=7 kpc.

According to the results of \citet{2010ApJ...716....1B}, for the disc
component, we approximate the velocity ellipsoid with principal axes
pointing along the directions of unitary vectors in cylindrical
coordinates.

\begin{table}
\caption{Parameters $a,b,c$ given by \citet{2010ApJ...716....1B} in the
 relation $a+b|z|^c$ for disc
 dispersions and mean disc velocity at $R_0$ and towards the north
Galactic Pole.}
\label{disco}
\begin{tabular}{cccc}
\hline
    & $a$ & $b$ & $c$ \\
\hline
 $\sigma_R(R_0,z)$ & 40 & 5 & 1.5 \\
 $\sigma_z(R_0,z)$ & 25 & 4 & 1.5 \\
 $\sigma_{\varphi}(R_0,z)$ & 30 & 3 & 2 \\
 <${v}_{\varphi}$>$(R_0,z)$ & $-$205 & 19.2 & 1.25 \\
\hline
\end{tabular}
\end{table}

\subsection{Bulge component}
\label{bulbo}

The Galactic Bulge has also several determinations of velocity
dispersion and mean rotation velocity, e.g. \citet{1988ApJ...325..563F,1992ApJ...387..181K,1995MNRAS.275..605I,
2000AJ....120..855B,2008ApJ...688.1060H,2010ApJ...720L..72S,
2012AJ....143...57K,2013MNRAS.432.2092N,2014A&A...562A..66Z,
2016ApJ...832..132Z,2018A&A...616A..83V,2020MNRAS.498.5629D,
2020AJ....159..270K,2021ApJ...908...21Z}. For this component, we take
 the results given by
\citet{2014A&A...562A..66Z}. They find some fits for the radial velocity
dispersion $\sigma$ and mean velocity $V$ in terms of Galactic
coordinates $(l,b)$. On the Galactic plane their fits in $\kms$ are

\begin{equation} \sigma= A_1+B_1+D_1l^2+E_1{\rm e}^{-l^2/s}, 
\label{sigb}
\end{equation}

\begin{equation} V=A_2+D_2\tanh(F_2l), 
\label{Vb}
\end{equation}

\noindent with $A_1$=79.39, $B_1$=38.45, $D_1$=$-$0.26, $E_1$=21.08,
$s$=2.47, and $A_2$=3.8, $D_2$=76.7, $F_2$=0.3; $l$ in degrees. These
values correspond to the projected distance $R_\perp=R_0\sin l$.
In our computations, we approximate as isotropic the dispersion field
of the bulge, based on values obtained with Eq.~\ref{sigb}. Thus,
 with $l=\frac{180}{\pi}\arcsin (r/R_0)$ and $r$
the distance from the Galactic Centre, in spherical coordinates we take

\begin{equation}  \sigma_r(r) =\sigma_{\varphi}(r)=\sigma_{\theta}(r)=118-0.26l^2+21{\rm e}^{-l^2/2.5},
\label{dispb}
\end{equation}

\noindent with $r \lesssim 0.363R_0$ to avoid negative values.

\citet{1995MNRAS.275..605I}, \citet{2012AJ....143...57K}, and
\citet{2013MNRAS.432.2092N} find that
the rotation in the bulge is approximately cylindrical.
Thus, with $l=\frac{180}{\pi}\arcsin (R/R_0)$ and $R$ the cylindrical
coordinate, we take

\begin{equation} <\!{v}_{\varphi}\!>(R)=-77\tanh(0.3l), 
\label{vrotb}
\end{equation}

\noindent here we have ignored the small term $A_2$ in Eq.~\ref{Vb},
and the minus sign gives a prograde rotation.

\subsection{Dark halo component}
\label{halo}

The Galactic Dark Halo has been studied in short and large distances
from the Sun, e.g. \citet{1994MNRAS.271...94S,1997ApJ...481..775S,
2000AJ....119.2843C,2005MNRAS.364..433B,
2006MNRAS.370.1055B,2006MNRAS.369.1688D,2009MNRAS.399.1223S,
2010ApJ...716....1B,2010ApJ...712..692C,2010A&A...510L...4S,
2010AJ....139...59B,2012ApJ...761...98K,2013MNRAS.432.2402F,
2014ApJ...794...59K,2015ApJ...813...89K,2017MNRAS.470.1259D,
2017MNRAS.464L..80P,2019AJ....157..104B,2019MNRAS.485.3296W}.
 Here, we consider motions only
within its inner region $r \lesssim$ 10 kpc. In this region,
\citet{2009MNRAS.399.1223S} and \citet{2010ApJ...716....1B} give
values for velocity dispersions and mean rotation velocity.
We take the following values from \citet{2010ApJ...716....1B} as 
Conditions I, with zero mean rotation:
$(\sigma_r,\sigma_{\varphi},\sigma_{\theta})$=(141,85,75) $\kms$,
<${v}_{\varphi}$>=0. 

For Conditions II, from figures 7 and 8 of \citet{2019MNRAS.485.3296W}
and within the region $R \lesssim$ 10 kpc, $|z| \lesssim$ 5 kpc,
we approximate the velocity dispersions and mean prograde rotation
 velocity with
$(\sigma_r,\sigma_{\varphi},\sigma_{\theta})$=(175,150,125) $\kms$,
<${v}_{\varphi}$>=$-$25 $\kms$. In these second
conditions the non-zero rotation approximates the values 
$-$37 $\kms$ given by \citet{2010A&A...510L...4S}, between
$-$50 and $-$30 $\kms$ of \citet{2000AJ....119.2843C}, and
between $-$5 and $-$25 $\kms$ in \citet{2017MNRAS.470.1259D}. 

In this dark halo component, we approximate the velocity ellipsoid with
principal axes pointing along the directions of unitary vectors in
spherical coordinates \citep{2009MNRAS.399.1223S,2010ApJ...716....1B}.

\subsection{The Galactic model}
\label{modgal}

The Galactic model employed in our computations is the axisymmetric
model of \citet{1991RMxAA..22..255A}. It has three components:
a Miyamoto-Nagai \citep{1975PASJ...27..533M} disc, a spherical bulge,
and a spherical dark halo. We rescaled this model to the
Sun's galactocentric distance $R_0$=8.15 kpc and Local Standard of Rest
velocity of 236 $\kms$, listed in table 3, column A5 of
\citet{2019ApJ...885..131R}. The Solar motion, also from this table,
is $(U,V,W)_{\odot}$=($-$10.6,\,10.7,\,7.6)$\kms$, with
$U_{\odot}$ negative towards the Galactic Centre.  
In this model, the computed orbits under the effect of dynamical
friction are obtained in an inertial reference frame with origin
at the Galactic Centre; the $x$-axis points to the present
position of the Sun, and the $y$-axis points in the opposite
direction to Galactic rotation. Prograde rotation velocity is 
in the sense of Galactic rotation, and has a negative sign.

The axisymmetric model of \citet{1991RMxAA..22..255A}
does not include the distinction between the thin and thick discs.
Future analyses of the dynamical friction effect will consider the
contributions of multiple components belonging to the thin and
thick discs, employing the more detailed Galactic
model GravPot16\footnote{\url{https://gravpot.utinam.cnrs.fr}} which
takes account of these components \citep{FT2017}. A more complete
analysis of this effect will be needed in the non-axisymmetric version
of this model, which includes a boxy bar and 3D spiral arms.

\section{Orbits of globular clusters}
\label{orbitas}

As stated in Section~\ref{dispvobs}, we only computed the motion of
globular clusters whose orbits lie within the Galactic region
$R \lesssim$ 10 kpc, $|z| \lesssim$ 5 kpc, with $R,z$ cylindrical
coordinates. With the data of globular clusters from the 
Holger Baumgardt compilation\footnote{\url{https://people.smp.uq.edu.au/HolgerBaumgardt/globular/}}, we made
a first computation of orbits employing the Galactic model in
Section~\ref{modgal} without the dynamical friction effect, and
separated the clusters satisfying the above condition.
Table~\ref{cumulos} gives the data for these clusters: position,
distance from the Sun, heliocentric radial velocity, proper motions,
mass, and half-mass radius, all these quantities taken from
the Holger Baumgardt compilation. In particular, the mass listed in
this table is the current mass of the clusters.

For some clusters in Table~\ref{cumulos}, and without the
dynamical friction effect,
in Appendix~\ref{ap4} we present a comparison of mean perigalactic
and apogalactic distances obtained with the Galactic model employed
in this work, and corresponding results from
\citet{2018A&A...616A..12G} and the Holger Baumgardt compilation,
obtained with Model I of \citet{2013A&A...549A.137I}, based on the
original Galactic model of \citet{1991RMxAA..22..255A}.
As shown in Table~\ref{cumulos.ax}, these perigalactic and apogalactic
distances compare well in these similar Galactic models; thus, our
results with the dynamical friction effect will represent the
orbital evolution under essentially an Allen \& Santill{\'a}n model.

The orbits of all the clusters in Table~\ref{cumulos} were computed
in a time interval of 5 Gyr in the future, considering the
effect of the dynamical friction. Galactic positions and velocities
computed with data in this table were taken as initial conditions in
the orbits. To test the accuracy of the numerical integration,
in some clusters we computed the orbits backwards in time and then
forwards in time up to the initial time $t$=0. The initial conditions
were recovered with good approximation. In Section~\ref{impl} we
show the orbits of some clusters computed backwards in time, i.e. in
the past.
As an approximation in all
our computations, during the orbital evolution, the mass of a cluster
was kept constant and equal to the current value listed in the
table. This overestimates the effect of dynamical
friction, as shown by Eq.~\ref{acfd}, because this mass will decrease
with time due to evaporation and tidal shocks, reinforced by
dynamical friction itself. Thus, the total
integration time of 5 Gyr is only a convenient time to see the trend
of the orbital evolution under the maximum dynamical friction effect.

For the computations, we employed the Runge--Kutta algorithm
 of seventh--eighth
order given by \citet{F68}. Besides finding the successive orbital
perigalactic and apogalactic distances $r_{\rm min}$, $r_{\rm max}$,
respectively, we computed at every
orbital point the instantaneous values per unit mass of the energy,
$E$, the $z$-component of the angular momentum, $L_z$, and their time
variations $\frac{{\rm d}E}{{\rm d}t}$, $\frac{{\rm d}L_z}{{\rm d}t}$.
With {\boldmath $V$} the velocity of the cluster with respect to
the Galactic inertial frame, and {\boldmath $a$}$_{\rm df}$ the total
acceleration due to dynamical friction, i.e. the second term on the
right side of Eq.~\ref{actot}, the time variations of $E$ and $L_z$
are given by

\begin{equation}
  \frac{{\rm d}E}{{\rm d}t}=\mbox{\boldmath $V$}\cdot \mbox{\boldmath $a$}_{\rm df},
\label{dEdt}
\end{equation}

\begin{equation}
  \frac{{\rm d}L_z}{{\rm d}t}=R(a_{\varphi})_{\rm df}, 
\label{dEdt}
\end{equation}

\noindent with $(a_{\varphi})_{\rm df}$ the azimutal component of
{\boldmath $a$}$_{\rm df}$.

\subsection{Mean variations}
\label{varprom}

\begin{figure}
\includegraphics[width=\columnwidth]{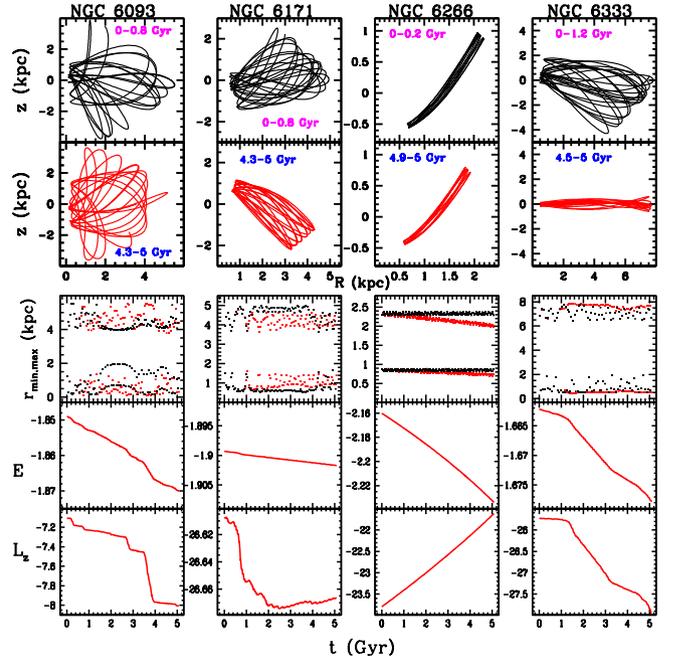}
\caption{Results for selected clusters. The panels in a column
correspond to the cluster with the name printed at the top.
From top to bottom, the first two panel rows show the
 meridional orbit,
in the interval of time given in the corresponding panel, without and
with the effect of dynamical friction, in black and red colours,
respectively. The third panel shows, as functions of time, the values of
$r_{\rm min}$ and $r_{\rm max}$, with (red colour) and without
(black colour) the effect of dynamical friction. The last
two panel rows give $E$, $L_z$ as functions of time,
 also with the effect
of dynamical friction. The units of $E$ and $L_z$ are
UE= 10$^5$ km$^2$ s$^{-2}$, UL = 10 kpc km s$^{-1}$.}
\label{fig1}
\end{figure}

\begin{figure}
\includegraphics[width=\columnwidth]{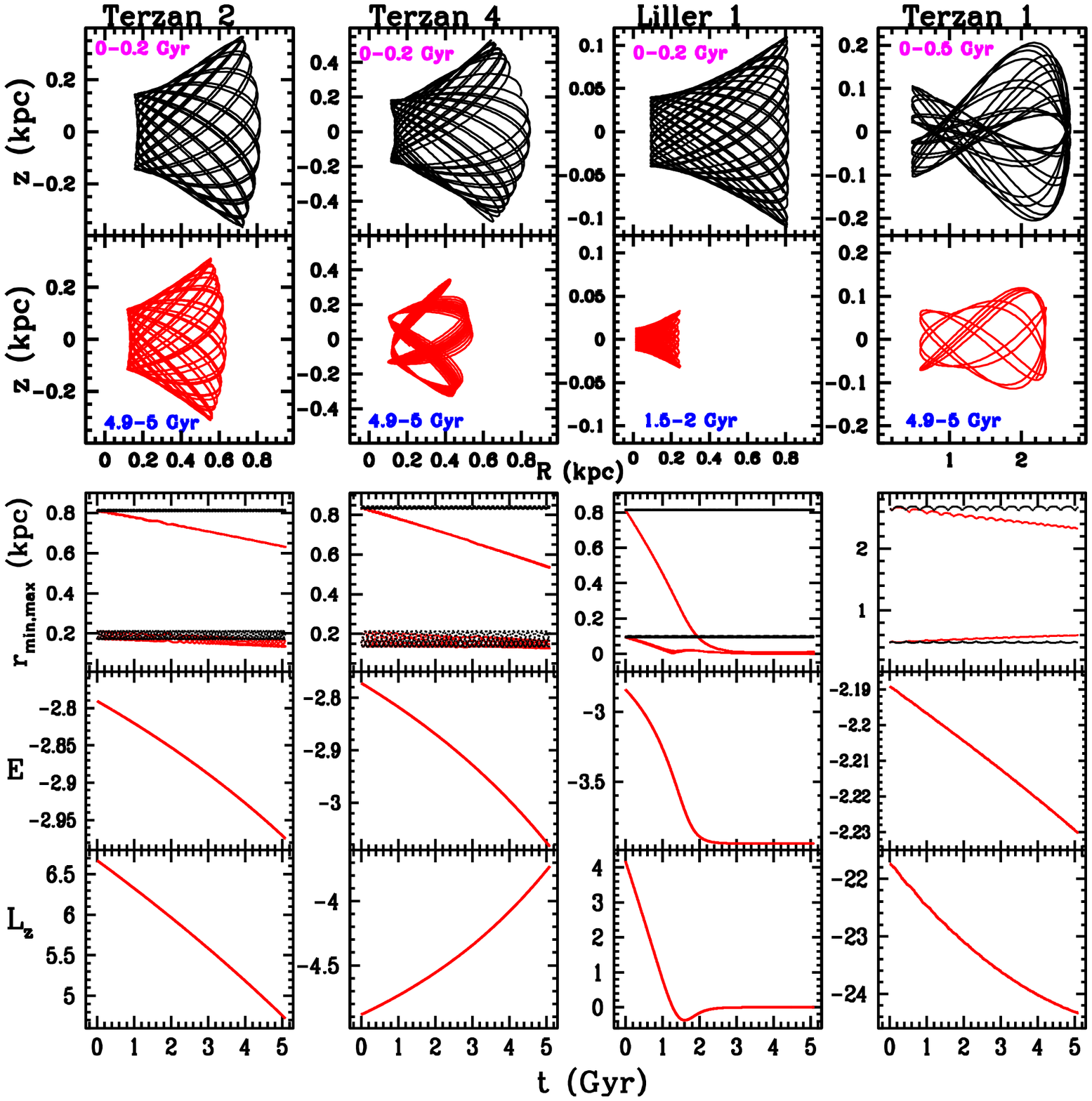}
\caption{Same as in Fig.~\ref{fig1}.}
\label{fig2}
\end{figure}

\begin{figure}
\includegraphics[width=\columnwidth]{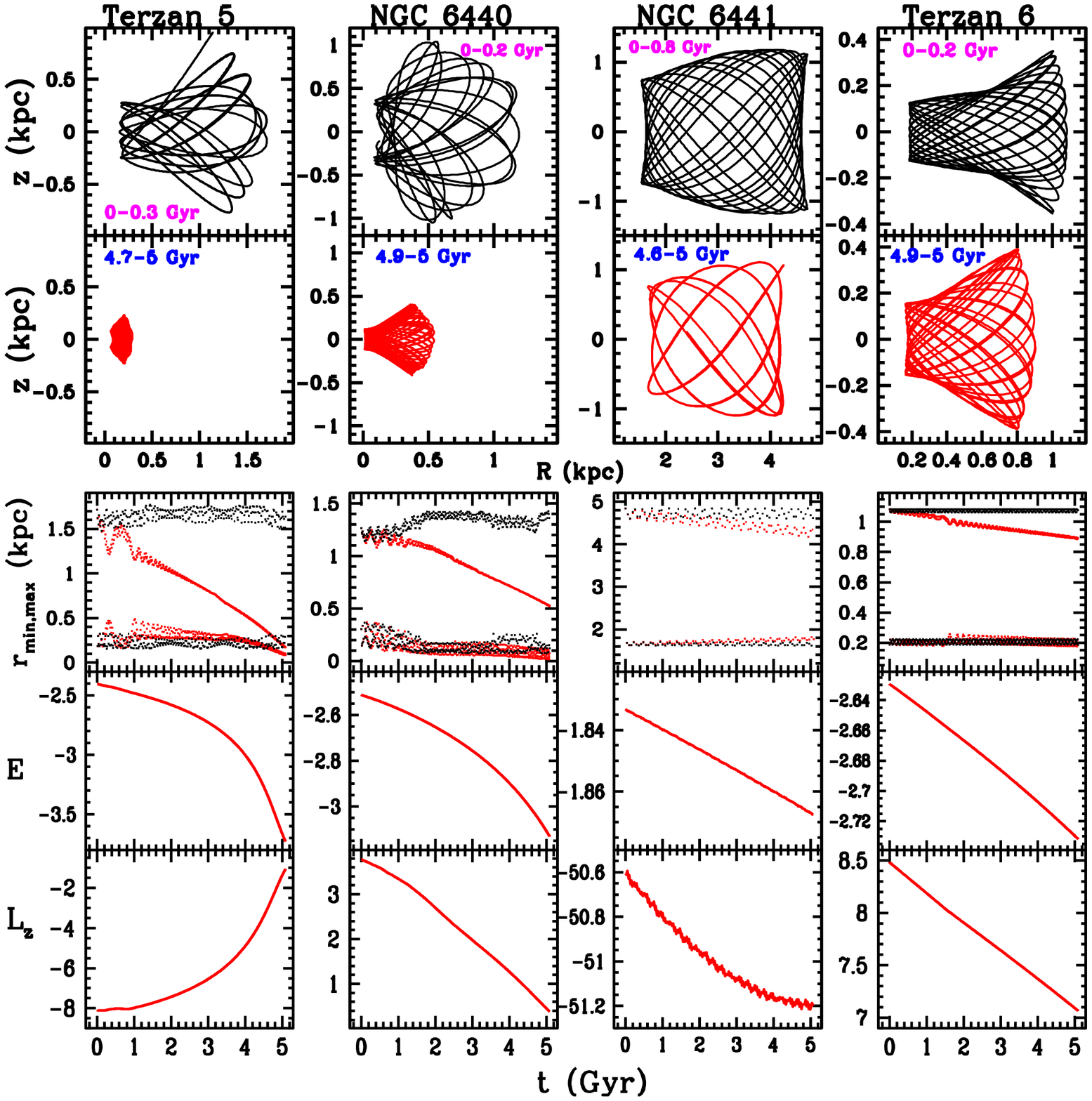}
\caption{Same as in Fig.~\ref{fig1}.}
\label{fig3}
\end{figure}

\begin{figure}
\includegraphics[width=\columnwidth]{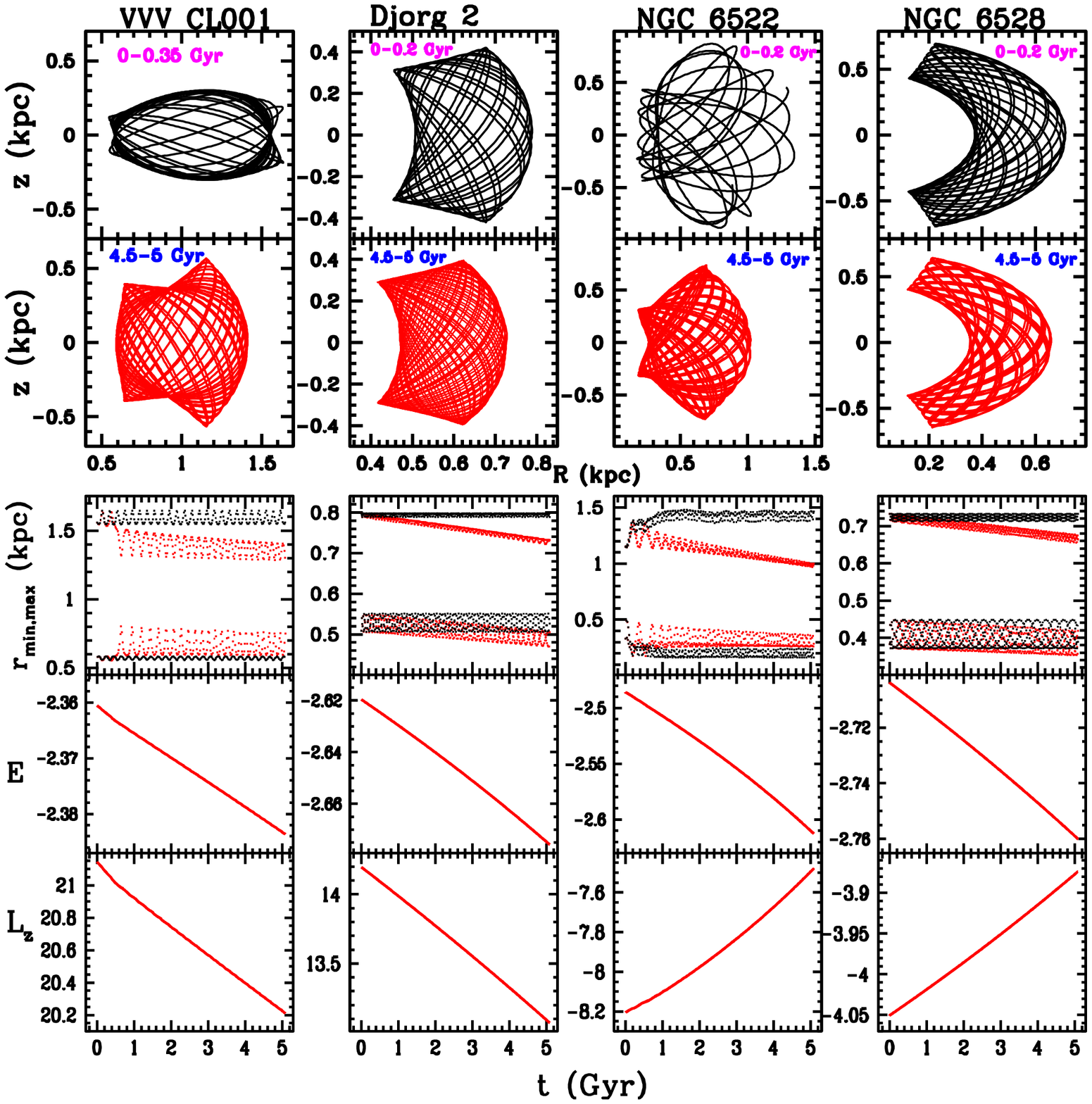}
\caption{Same as in Fig.~\ref{fig1}.}
\label{fig4}
\end{figure}

\begin{figure}
\includegraphics[width=\columnwidth]{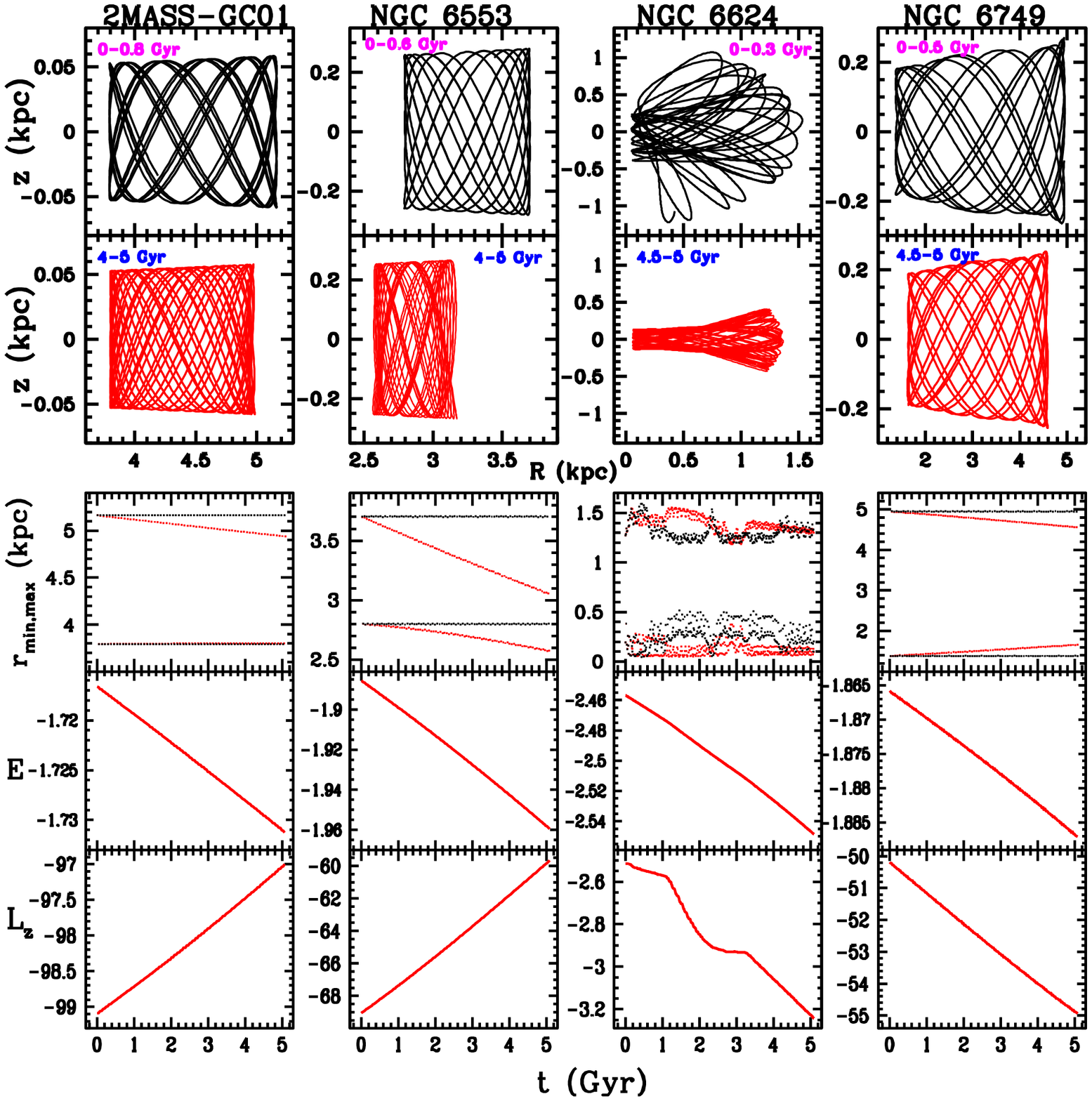}
\caption{Same as in Fig.~\ref{fig1}.}
\label{fig5}
\end{figure}

As examples, Figs.~\ref{fig1}--\ref{fig5} show some orbital properties
obtained in selected clusters.
We present results mainly using Conditions I in the Galactic
dark halo (see Section~\ref{halo}); Conditions II give similar
results. All the panels in a column correspond to the cluster with the
name printed at the top. From top to bottom, the first two panel
rows show
the meridional orbit, without and with the effect of dynamical
friction, respectively in black and red colours. $R$ is
distance in cylindrical coordinates. Each orbit is shown in
the interval of time given in the corresponding panel; the orbits in
red colour are shown in advanced times near 5 Gyr. Details of the
orbital evolution are given in the remaining three panels:
the third panel row shows as functions of time the values of
$r_{\rm min}$ and $r_{\rm max}$, with (red colour) and without 
(black colour) the effect of dynamical friction. $r$ is
distance in spherical coordinates. The last two panel rows
give $E$, $L_z$ as functions of time, also with the effect
of dynamical friction. In these figures the units of $E$ and $L_z$ are
UE= 10$^5$ km$^2$ s$^{-2}$, UL = 10 kpc km s$^{-1}$.

Several orbits in Figs.~\ref{fig1}--\ref{fig5} show strong variations
in $r_{\rm min}$ and $r_{\rm max}$, particularly the orbit of 
Liller 1 in Fig.~\ref{fig2}, which falls into the 
Galactic nuclear region in about 2--3 Gyr. Also, in some clusters the 
mean time-variation of $L_z$ is positive, and negative in others.
At initial times the orbits of some clusters with and without
the effect of dynamical friction can be similar, but they begin to
deviate at later times. This behaviour reflects the slight initial
effect of dynamical friction in these cases.

In clusters like NGC 6093 and NGC 6171 in Fig.~\ref{fig1} the trend
of the mean time-variation of $r_{\rm min}$ and $r_{\rm max}$
is not clearly defined. Thus, we separated the clusters in
which these perigalactic and apogalactic time variations could be
obtained with reasonable certainty, for example, NGC 6266 in
Fig.~\ref{fig1}. In this separated sample, we find that
$r_{\rm min}$ and $r_{\rm max}$ vary approximately linearly with
time over the entire 5 Gyr interval, as shown in some of
its members in Figs.~\ref{fig1}--\ref{fig5}. 
Thus, with linear least-square fits over this interval we estimated 
$<\!\dot{r}_{\rm min}\!>$, $<\!\dot{r}_{\rm max}\!>$. In 
Liller 1 these fits were done only in the first Gyr, due to its
rapid later change.

As a result of this linear behaviour, the obtained
$<\!\dot{r}_{\rm min}\!>$, $<\!\dot{r}_{\rm max}\!>$ can be applied
directly in the first Gyr or early times in the computations, where
the assumption of constant mass of the clusters is appropriate.
Also, in the separated cluster sample, and in line with the assumption
of constant mass, the mean time-variations of energy and angular 
momentum, $<\!\dot{E}\!>$, $<\!\dot{L}_z\!>$, were estimated only over
the first Gyr with least-square fits; energy and angular momentum 
can have different variations in extended time intervals, as shown in
Figs.~\ref{fig1}--\ref{fig5}.

The results of this analysis are listed in Table~\ref{prop1}.
The initial values $r_{\rm min_0}$, $r_{\rm max_0}$ are the first
perigalactic and apogalactic distances obtained at, or after the start,
$t$=0, of an orbital computation, and $E_0$, $L_{z_0}$ are
directly obtained at $t$=0. The last column
lists the associated Main Progenitor, given by
\citet{2019A&A...630L...4M}: cluster formed in situ in the
disc (M-D), bulge (M-B), unassociated low energy (L-E),
and coming from an accretion event: $Gaia$-Enceladus (G-E),
Sequoia (Seq).

Strong variations in $r_{\rm max}$ are obtained in the inner
Galaxy clusters Liller 1, Terzan 5, NGC 6440, and NGC 6553.
The first three clusters have an initial angular
momentum $L_{z_0}$ of low magnitude. Terzan 2 and Terzan 4,
also with low $L_{z_0}$, have a slightly smaller variation in
$r_{\rm max}$. NGC 5139 (Omega Centauri) has moderate to
strong variations in $r_{\rm min}$, $r_{\rm max}$.

The values of $<\!\dot{r}_{\rm min}\!>$,
$<\!\dot{r}_{\rm max}\!>$ in Table~\ref{prop1} are in units of
pc Gyr$^{-1}$, which in some clusters can be quite small, mainly in
$<\!\dot{r}_{\rm min}\!>$. Their corresponding listed uncertainties
can be of the same order of the mean values, but even without the
effect of dynamical friction, and so under $E$, $L_z$ constant,
some clusters can present strong variations of $r_{\rm min}$,
$r_{\rm max}$, e.g. NGC 6093 in Fig.~\ref{fig1}.
The uncertainties in the mean time-variations were estimated
computing the orbits with initial minimum and maximum energies in each
cluster, considering the effect of dynamical friction, according to the
uncertainties in distance, radial velocity, and proper motions listed
in Table~\ref{cumulos};
\citet{2014ApJ...793..110M} find that this procedure gives
uncertainty estimates that approximates those obtained with a Monte
Carlo simulation. 

In the clusters not included in Table~\ref{prop1}, we computed only 
the mean time-variations $<\!\dot{E}\!>$, $<\!\dot{L}_z\!>$ by means
of linear least-squares fits, taking the
time interval of the first Gyr. These variations, along with
$r_{\rm min_0}$, $r_{\rm max_0}$, $E_0$, $L_{z_0}$ are listed in
Table~\ref{prop2}. The symbols indicating progenitors shown in
this table are as in Table~\ref{prop1}; the new symbol K stands for the
$Koala$ dwarf galaxy accretion event \citep{2020MNRAS.493..847F}.
In the old, metal-deficient, cluster VVV CL001, not listed by
\citet{2019A&A...630L...4M} and \citet{2020MNRAS.493..847F}, a possible association
with Sequoia or Gaia-Enceladus structures has been suggested by
\citet{2021ApJ...908L..42F}, and this is listed in Table~\ref{prop2}.
Fig.~\ref{fig4} shows a moderate effect of dynamical friction in this
cluster. Also, in a recent study, based on the age-metallicity
relation, Pal 6 is shown to be connected with the M-B group
\citep{2021arXiv210904483S}, instead of the L-E group proposed by 
\citet{2019A&A...630L...4M}. For the strongly affected clusters
 Terzan 4, Terzan 5, 
NGC6440, and NGC 6553, \citet{2019A&A...630L...4M} provide an in situ
bulge origin, but the extreme effect of dynamical friction in these
cases requires a future detailed analysis of these clusters,
including Liller 1 with uncertain classification by
\citeauthor{2019A&A...630L...4M}.

\begin{table*}
 \caption{Globular clusters with orbits lying in the Galactic region
$R \lesssim$ 10 kpc, $|z| \lesssim$ 5 kpc.}
 \label{cumulos}
 \begin{tabular}{ccccccccc}
  \hline
 Cluster & $\alpha$ & $\delta$ & $r$ & $v_r$ & $\mu_x$ & $\mu_y$ &
 $M_c$ & $r_h$  \\
         & (deg) & (deg) & (kpc) & ($\kms$) & (mas yr$^{-1}$) & (mas yr$^{-1}$) & ($M_{\sun}$) & (pc)  \\
  \hline
 NGC 104 & 6.023792 & $-$72.081306 &  4.52$\pm$0.03 & $-$17.45$\pm$0.16 &  5.252$\pm$0.021 & $-$2.551$\pm$0.021 & 8.95$\times 10^5$ & 6.30 \\ 
 NGC 4372 & 186.439101 & $-$72.659084 &  5.71$\pm$0.21 & 75.59$\pm$0.30 & $-$6.409$\pm$0.024 & 3.297$\pm$0.024 & 1.98$\times 10^5$ & 8.53 \\ 
 BH 140 & 193.472915 & $-$67.177276 & 4.81$\pm$0.25 & 90.30$\pm$0.35 &  $-$14.848$\pm$0.024 & 1.224$\pm$0.024 & 5.99$\times 10^4$ & 9.53  \\ 
 NGC 4833 & 194.891342 & $-$70.876503 & 6.48$\pm$0.08 & 201.99$\pm$0.40 & $-$8.377$\pm$0.025 & $-$0.963$\pm$0.025 & 2.06$\times 10^5$ & 4.76 \\
 NGC 5139 & 201.696991 & $-$47.479473 & 5.43$\pm$0.05 & 232.78$\pm$0.21 & $-$3.341$\pm$0.028 & $-$6.557$\pm$0.043 & 3.64$\times 10^6$ & 10.36 \\
 NGC 5927 & 232.002869 & $-$50.673031 & 8.27$\pm$0.11 & $-$104.09$\pm$0.28 & $-$5.056$\pm$0.025 & $-$3.217$\pm$0.025 & 2.75$\times 10^5$ & 5.28 \\ 
 NGC 5946 & 233.869051 & $-$50.659713 & 9.64$\pm$0.51 & 137.60$\pm$0.94 & $-$5.331$\pm$0.028 & $-$1.657$\pm$0.027 & 9.31$\times 10^4$ & 2.59 \\ 
 NGC 5986 & 236.512497 & $-$37.786415 & 10.54$\pm$0.13 & 101.18$\pm$0.43 & $-$4.192$\pm$0.026 & $-$4.568$\pm$0.026 & 3.34$\times 10^5$ & 4.25 \\
 FSR 1716 & 242.625000 & $-$53.748889 & 7.43$\pm$0.27 & $-$30.70$\pm$0.98 & $-$4.354$\pm$0.033 & $-$8.832$\pm$0.031 & 6.43$\times 10^4$ & 5.16\\
 Lynga 7 & 242.765213 & $-$55.317776 & 7.90$\pm$0.16 & 17.86$\pm$0.83 &  $-$3.851$\pm$0.027 & $-$7.050$\pm$0.027 & 7.96$\times 10^4$ & 5.16  \\ 
 NGC 6093 & 244.260040 & $-$22.976084 & 10.34$\pm$0.12 & 10.93$\pm$0.39 & $-$2.934$\pm$0.027 & $-$5.578$\pm$0.026 & 3.38$\times 10^5$ & 2.62 \\ 
 NGC 6121 & 245.896744 & $-$26.525749 & 1.85$\pm$0.02 & 71.21$\pm$0.15 & $-$12.514$\pm$0.023 & $-$19.022$\pm$0.023 & 8.71$\times 10^4$ & 3.69 \\
 NGC 6144 & 246.807770 & $-$26.023500 & 8.15$\pm$0.13 & 194.79$\pm$0.58 & $-$1.744$\pm$0.026 & $-$2.607$\pm$0.026 & 7.92$\times 10^4$ & 4.91 \\ 
 NGC 6139 & 246.918466 & $-$38.848782 & 10.04$\pm$0.45 & 24.41$\pm$0.95 & $-$6.081$\pm$0.027 & $-$2.711$\pm$0.026 & 3.23$\times 10^5$ & 2.47 \\ 
 Terzan 3 & 247.162483 & $-$35.339829 & 7.64$\pm$0.31 & $-$135.76$\pm$0.57 & $-$5.577$\pm$0.027 & $-$1.760$\pm$0.026 & 4.04$\times 10^4$ & 7.19 \\ 
 NGC 6171 & 248.132752 & $-$13.053778 & 5.63$\pm$0.08 & $-$34.71$\pm$0.18 & $-$1.939$\pm$0.025 & $-$5.979$\pm$0.025 & 7.49$\times 10^4$ & 3.94\\
 ESO 452-SC11 & 249.854167 & $-$28.399167 & 7.39$\pm$0.20 & 16.37$\pm$0.44 & $-$1.423$\pm$0.031 & $-$6.472$\pm$0.030 & 8.26$\times 10^3$ & 3.68 \\ 
 NGC 6218 & 251.809067 & $-$1.948528 & 5.11$\pm$0.05 & $-$41.67$\pm$0.14 & $-$0.191$\pm$0.024 & $-$6.802$\pm$0.024 & 1.07$\times 10^5$ & 4.05 \\
 FSR 1735 & 253.044174 & $-$47.058056 & 9.08$\pm$0.53 & $-$69.85$\pm$4.88 & $-$4.439$\pm$0.054 & $-$1.534$\pm$0.048 & 7.23$\times 10^4$ & 2.97\\
 NGC 6235 & 253.355676 & $-$22.177447 & 11.94$\pm$0.38 & 126.68$\pm$0.33 & $-$3.931$\pm$0.027 & $-$7.587$\pm$0.027 & 1.07$\times 10^5$ & 4.78 \\
 NGC 6254 & 254.287720 & $-$4.100306 & 5.07$\pm$0.06 & 74.21$\pm$0.23 &  $-$4.758$\pm$0.024 & $-$6.597$\pm$0.024 & 2.05$\times 10^5$ & 4.81 \\ 
 NGC 6256 & 254.886107 & $-$37.120968 & 7.24$\pm$0.29 & $-$99.75$\pm$0.66 & $-$3.715$\pm$0.031 & $-$1.637$\pm$0.030 & 1.25$\times 10^5$ & 4.82\\
 NGC 6266 & 255.304153 & $-$30.113390 & 6.41$\pm$0.10 & $-$73.98$\pm$0.67 & $-$4.978$\pm$0.026 & $-$2.947$\pm$0.026 & 6.10$\times 10^5$ & 2.43\\
 NGC 6273 & 255.657486 & $-$26.267971 & 8.34$\pm$0.16 & 145.54$\pm$0.59 & $-$3.249$\pm$0.026 & 1.660$\pm$0.025 & 6.97$\times 10^5$ & 4.21 \\ 
 NGC 6284 & 256.120114 & $-$24.764799 & 14.21$\pm$0.42 & 28.62$\pm$0.73 & $-$3.200$\pm$0.029 & $-$2.002$\pm$0.028 & 1.29$\times 10^5$ & 3.78 \\ 
 NGC 6287 & 256.288904 & $-$22.708005 & 7.93$\pm$0.37 & $-$294.74$\pm$1.65 & $-$5.010$\pm$0.029 & $-$1.883$\pm$0.028 & 1.45$\times 10^5$ & 3.65\\
 NGC 6293 & 257.542500 & $-$26.582083 & 9.19$\pm$0.28 & $-$143.66$\pm$0.39 & 0.870$\pm$0.028 & $-$4.326$\pm$0.028 & 2.05$\times 10^5$ & 4.05 \\ 
 NGC 6304 & 258.634399 & $-$29.462028 & 6.15$\pm$0.15 & $-$108.62$\pm$0.39 & $-$4.070$\pm$0.029 & $-$1.088$\pm$0.028 & 1.26$\times 10^5$ & 4.26  \\ 
 NGC 6316 & 259.155417 & $-$28.140111 & 11.15$\pm$0.39 & 99.65$\pm$0.84 & $-$4.969$\pm$0.031 & $-$4.592$\pm$0.030 & 3.18$\times 10^5$ & 4.77 \\ 
 NGC 6325 & 259.496327 & $-$23.767677 & 7.53$\pm$0.32 & 29.54$\pm$0.58 & $-$8.289$\pm$0.030 & $-$9.000$\pm$0.029 & 5.89$\times 10^4$ & 2.05  \\ 
 NGC 6333 & 259.799086 & $-$18.516257 & 8.30$\pm$0.14 & 310.75$\pm$2.12 & $-$2.180$\pm$0.026 & $-$3.222$\pm$0.026 & 3.23$\times 10^5$ & 4.17 \\ 
 NGC 6342 & 260.291573 & $-$19.587659 & 8.01$\pm$0.23 & 115.75$\pm$0.90 & $-$2.903$\pm$0.027 & $-$7.116$\pm$0.026 & 4.22$\times 10^4$ & 2.06 \\ 
 NGC 6356 & 260.895804 & $-$17.813027 & 15.66$\pm$0.92 & 48.18$\pm$1.82 & $-$3.750$\pm$0.026 & $-$3.392$\pm$0.026 & 6.00$\times 10^5$ & 6.86 \\ 
 NGC 6355 & 260.993533 & $-$26.352827 & 8.65$\pm$0.22 & $-$195.85$\pm$0.55 & $-$4.738$\pm$0.031 & $-$0.572$\pm$0.030 & 1.01$\times 10^5$ & 3.55  \\ 
 NGC 6352 & 261.371277 & $-$48.422169 & 5.54$\pm$0.07 & $-$125.63$\pm$1.01 & $-$2.158$\pm$0.025 & $-$4.447$\pm$0.025 & 6.47$\times 10^4$ & 4.56  \\ 
 Terzan 2 & 261.887917 & $-$30.802333 & 7.75$\pm$0.33 & 134.56$\pm$0.96 & $-$2.170$\pm$0.041 & $-$6.263$\pm$0.038 & 1.36$\times 10^5$ & 4.16 \\ 
 NGC 6366 & 261.934357 & $-$5.079861 & 3.44$\pm$0.05 & $-$120.65$\pm$0.19 & $-$0.332$\pm$0.025 & $-$5.160$\pm$0.024 & 3.76$\times 10^4$ & 5.56\\
 Terzan 4 & 262.662506 & $-$31.595528 & 7.59$\pm$0.31 & $-$48.96$\pm$1.57 & $-$5.462$\pm$0.060 & $-$3.711$\pm$0.048 & 2.00$\times 10^5$ & 6.06\\
 HP 1 & 262.771667 & $-$29.981667 & 7.00$\pm$0.14 & 39.76$\pm$1.22 &    2.523$\pm$0.039 & $-$10.093$\pm$0.037 & 1.24$\times 10^5$ & 3.74  \\ 
 NGC 6362 & 262.979096 & $-$67.048332 & 7.65$\pm$0.07 & $-$14.58$\pm$0.18 & $-$5.506$\pm$0.024 & $-$4.763$\pm$0.024 & 1.27$\times 10^5$ & 7.23\\
 Liller 1 & 263.352333 & $-$33.389556 & 8.06$\pm$0.34 & 60.36$\pm$2.44 &  $-$5.403$\pm$0.109 & $-$7.431$\pm$0.077 & 9.15$\times 10^5$ & 2.01 \\ 
 NGC 6380 & 263.618611 & $-$39.069530 & 9.61$\pm$0.30 & $-$1.48$\pm$0.73 & $-$2.183$\pm$0.031 & $-$3.233$\pm$0.030 & 3.34$\times 10^5$ & 4.40 \\
 Terzan 1 & 263.946667 & $-$30.481778 & 5.67$\pm$0.17 & 56.75$\pm$1.61 & $-$2.806$\pm$0.055 & $-$4.861$\pm$0.055 & 1.50$\times 10^5$ & 2.15  \\ 
 Ton 2 & 264.039293 & $-$38.540925 & 6.99$\pm$0.34 & $-$184.72$\pm$1.12 & $-$5.904$\pm$0.031 & $-$0.755$\pm$0.029 & 6.91$\times 10^4$ & 4.60 \\ 
 NGC 6388 & 264.071777 & $-$44.735500 & 11.17$\pm$0.16 & 83.11$\pm$0.45 & $-$1.316$\pm$0.026 & $-$2.709$\pm$0.026 & 1.25$\times 10^6$ & 4.34 \\ 
 NGC 6402 & 264.400651 & $-$3.245916 & 9.14$\pm$0.25 & $-$60.71$\pm$0.45 & $-$3.590$\pm$0.025 & $-$5.059$\pm$0.025 & 5.92$\times 10^5$ & 5.14 \\
 NGC 6401 & 264.652191 & $-$23.909605 & 8.06$\pm$0.24 & $-$105.44$\pm$2.50 & $-$2.748$\pm$0.035 & 1.444$\pm$0.034 & 1.45$\times 10^5$ & 3.28 \\ 
 NGC 6397 & 265.175385 & $-$53.674335 & 2.48$\pm$0.02 & 18.51$\pm$0.08 &  3.260$\pm$0.023 & $-$17.664$\pm$0.022 & 9.66$\times 10^4$ & 3.90 \\ 
 Pal 6 & 265.925812 & $-$26.224995 & 7.05$\pm$0.45 & 177.00$\pm$1.35 &  $-$9.222$\pm$0.038 & $-$5.347$\pm$0.036 & 9.45$\times 10^4$ & 2.89 \\ 
  Djorg 1 & 266.869583 & $-$33.066389 & 9.88$\pm$0.65 & $-$359.18$\pm$1.64 & $-$4.693$\pm$0.046 & $-$8.468$\pm$0.041 & 7.97$\times 10^4$ & 5.57  \\ 
 Terzan 5 & 267.020200 & $-$24.779055 & 6.62$\pm$0.15 & $-$82.57$\pm$0.73 & $-$1.989$\pm$0.068 & $-$5.243$\pm$0.066 & 9.35$\times 10^5$ & 3.77\\
 NGC 6440 & 267.220167 & $-$20.360417 & 8.25$\pm$0.24 & $-$69.39$\pm$0.93 & $-$1.187$\pm$0.036 & $-$4.020$\pm$0.035 & 4.89$\times 10^5$ & 2.14\\
 NGC 6441 & 267.554413 & $-$37.051445 & 12.73$\pm$0.16 & 18.47$\pm$0.56 & $-$2.551$\pm$0.028 & $-$5.348$\pm$0.028 & 1.32$\times 10^6$ & 3.47 \\ 
 Terzan 6 & 267.693250 & $-$31.275389 & 7.27$\pm$0.35 & 136.45$\pm$1.50 & $-$4.979$\pm$0.048 & $-$7.431$\pm$0.039 & 1.04$\times 10^5$ & 1.33 \\ 
 NGC 6453 & 267.715508 & $-$34.598477 & 10.07$\pm$0.22 & $-$99.23$\pm$1.24 & 0.203$\pm$0.036 & $-$5.934$\pm$0.037 & 1.65$\times 10^5$ & 3.85 \\ 
 UKS 1 & 268.613312 & $-$24.145277 & 15.58$\pm$0.56 & 59.38$\pm$2.63 & $-$2.040$\pm$0.095 & $-$2.754$\pm$0.063 & 7.70$\times 10^4$ & 3.84 \\  
 VVV CL001 & 268.677083 & $-$24.014722 & 8.08$\pm$1.48 & $-$327.28$\pm$0.90 & $-$3.487$\pm$0.144 & $-$1.652$\pm$0.107 & 1.35$\times 10^5$ & 2.94  \\ 
 NGC 6496 & 269.765350 & $-$44.265945 & 9.64$\pm$0.15 & $-$134.72$\pm$0.26 & $-$3.060$\pm$0.027 & $-$9.271$\pm$0.026 & 6.89$\times 10^4$ & 6.42  \\ 
 Terzan 9 & 270.411667 & $-$26.839722 & 5.77$\pm$0.34 & 68.49$\pm$0.56 & $-$2.121$\pm$0.052 & $-$7.763$\pm$0.049 & 1.20$\times 10^5$ & 1.90  \\ 
 Djorg 2 & 270.454378 & $-$27.825819 & 8.76$\pm$0.18 & $-$149.75$\pm$1.10 & 0.662$\pm$0.042 & $-$2.983$\pm$0.037 & 1.25$\times 10^5$ & 5.16  \\ 
 NGC 6517 & 270.460750 & $-$8.958778 & 9.23$\pm$0.56 & $-$35.06$\pm$1.65 & $-$1.551$\pm$0.029 & $-$4.470$\pm$0.028 & 1.95$\times 10^5$ & 2.29 \\
 Terzan 10 & 270.740833 & $-$26.066944 & 10.21$\pm$0.40 & 211.37$\pm$2.27 & $-$6.827$\pm$0.059 & $-$2.588$\pm$0.050 & 3.02$\times 10^5$ & 4.60\\
  \hline
 \end{tabular}
\end{table*}

\begin{table*}
 \contcaption{}
 \label{cumulos.cont}
 \begin{tabular}{ccccccccc}
  \hline
 Cluster & $\alpha$ & $\delta$ & $r$ & $v_r$ & $\mu_x$ & $\mu_y$ &
 $M_c$ & $r_h$  \\
         & (deg) & (deg) & (kpc) & ($\kms$) & (mas yr$^{-1}$) & (mas yr$^{-1}$) & ($M_{\sun}$) & (pc)  \\
  \hline
 NGC 6522 & 270.891977 & $-$30.033974 & 7.29$\pm$0.21 & $-$15.23$\pm$0.49 & 2.566$\pm$0.039 & $-$6.438$\pm$0.036 & 2.11$\times 10^5$ & 3.08  \\ 
 NGC 6535 & 270.960449 & $-$0.297639 & 6.36$\pm$0.12 & $-$214.85$\pm$0.46 & $-$4.214$\pm$0.027 & $-$2.939$\pm$0.026 & 2.19$\times 10^4$ & 3.65\\
 NGC 6528 & 271.206697 & $-$30.055778 & 7.83$\pm$0.24 & 211.86$\pm$0.43 & $-$2.157$\pm$0.043 & $-$5.649$\pm$0.039 & 5.67$\times 10^4$ & 2.73 \\ 
 NGC 6539 & 271.207276 & $-$7.585858 & 8.16$\pm$0.39 & 35.19$\pm$0.50 &  $-$6.896$\pm$0.026 & $-$3.537$\pm$0.026 & 2.09$\times 10^5$ & 5.18  \\ 
 NGC 6540 & 271.535657 & $-$27.765286 & 5.91$\pm$0.27 & $-$16.50$\pm$0.78 & $-$3.702$\pm$0.032 & $-$2.791$\pm$0.032 & 3.45$\times 10^4$ & 5.32\\
 NGC 6544 & 271.833833 & $-$24.998222 & 2.58$\pm$0.06 & $-$38.46$\pm$0.67 & $-$2.304$\pm$0.031 & $-$18.604$\pm$0.030 & 9.14$\times 10^4$ & 2.07 \\ 
 NGC 6541 & 272.009827 & $-$43.714889 & 7.61$\pm$0.10 & $-$163.97$\pm$0.46 & 0.287$\pm$0.025 & $-$8.847$\pm$0.025 & 2.93$\times 10^5$ & 4.34 \\ 
 2MASS-GC01 & 272.090851 & $-$19.829723 & 3.37$\pm$0.62 & $-$31.28$\pm$0.50 & $-$1.121$\pm$0.296 & $-$1.881$\pm$0.235 & 3.50$\times 10^4$ & 4.70 \\ 
 NGC 6553 & 272.315322 & $-$25.907750 & 5.33$\pm$0.13 & $-$0.27$\pm$0.34 & 0.344$\pm$0.030 & $-$0.454$\pm$0.029 & 2.85$\times 10^5$ & 4.56  \\ 
 2MASS-GC02 & 272.402100 & $-$20.778889 & 5.50$\pm$0.44 & $-$237.75 $\pm$10.10 & 4.000$\pm$0.900 & $-$4.700$\pm$0.800 & 1.60$\times 10^4$ & 2.85  \\ 
 NGC 6558 & 272.573974 & $-$31.764508 & 7.47$\pm$0.29 & $-$195.12$\pm$0.73 & $-$1.720$\pm$0.036 & $-$4.144$\pm$0.034 & 2.65$\times 10^4$ & 1.70  \\ 
 IC 1276 & 272.684441 & $-$7.207595 & 4.55$\pm$0.25 & 155.06$\pm$0.69 &   $-$2.553$\pm$0.026 & $-$4.568$\pm$0.026 & 7.39$\times 10^4$ & 5.21 \\ 
 Terzan 12 & 273.065833 & $-$22.741944 & 5.17$\pm$0.38 & 95.61$\pm$1.21 & $-$6.222$\pm$0.037 & $-$3.052$\pm$0.034 & 8.72$\times 10^4$ & 3.28 \\ 
 NGC 6569 & 273.411667 & $-$31.826889 & 10.53$\pm$0.26 & $-$49.83$\pm$0.50 & $-$4.125$\pm$0.028 & $-$7.354$\pm$0.028 & 2.36$\times 10^5$ & 3.85 \\ 
 BH 261 & 273.527500 & $-$28.635000 & 6.12$\pm$0.26 & $-$45.00$\pm$15.00 & 3.566$\pm$0.043 & $-$3.590$\pm$0.037 & 2.20$\times 10^4$ & 4.66 \\ 
 NGC 6624 & 275.918793 & $-$30.361029 & 8.02$\pm$0.11 & 54.79$\pm$0.40 &   0.124$\pm$0.029 & $-$6.936$\pm$0.029 & 1.56$\times 10^5$ & 3.69 \\ 
 NGC 6626 & 276.137039 & $-$24.869847 & 5.37$\pm$0.10 & 11.11$\pm$0.60 & $-$0.278$\pm$0.028 & $-$8.922$\pm$0.028 & 2.99$\times 10^5$ & 2.26  \\ 
 NGC 6638 & 277.733734 & $-$25.497473 & 9.78$\pm$0.34 & 8.63$\pm$2.00 &  $-$2.518$\pm$0.029 & $-$4.076$\pm$0.029 & 1.18$\times 10^5$ & 2.20  \\ 
 NGC 6637 & 277.846252 & $-$32.348084 & 8.90$\pm$0.10 & 47.48$\pm$1.00 & $-$5.034$\pm$0.028 & $-$5.832$\pm$0.028 & 1.55$\times 10^5$ & 3.69  \\ 
 NGC 6642 & 277.975957 & $-$23.475602 & 8.05$\pm$0.20 & $-$60.61$\pm$1.35 & $-$0.173$\pm$0.030 & $-$3.892$\pm$0.030 & 3.44$\times 10^4$ & 1.51\\
 NGC 6652 & 278.940125 & $-$32.990723 & 9.46$\pm$0.14 & $-$95.37$\pm$0.86 & $-$5.484$\pm$0.027 & $-$4.274$\pm$0.027 & 4.81$\times 10^4$ & 1.96\\
 NGC 6656 & 279.099762 & $-$23.904749 & 3.30$\pm$0.04 & $-$148.72$\pm$0.78 & 9.851$\pm$0.023 & $-$5.617$\pm$0.023 & 4.76$\times 10^5$ & 5.29 \\ 
 Pal 8 & 280.377290 & $-$19.828858 & 11.32$\pm$0.63 & $-$31.54$\pm$0.21 & $-$1.987$\pm$0.027 & $-$5.694$\pm$0.027 & 6.74$\times 10^4$ & 5.86 \\ 
 NGC 6681 & 280.803162 & $-$32.292110 & 9.36$\pm$0.11 & 216.62$\pm$0.84 &  1.431$\pm$0.027 & $-$4.744$\pm$0.026 & 1.16$\times 10^5$ & 2.89 \\ 
 NGC 6712 & 283.268021 & $-$8.705960 & 7.38$\pm$0.24 & $-$107.45$\pm$0.29 & 3.363$\pm$0.027 & $-$4.436$\pm$0.027 & 9.63$\times 10^4$ & 3.21  \\ 
 NGC 6717 & 283.775177 & $-$22.701473 & 7.52$\pm$0.13 & 30.25$\pm$0.90 & $-$3.125$\pm$0.027 & $-$5.008$\pm$0.027 & 3.58$\times 10^4$ & 4.23  \\ 
 NGC 6723 & 284.888123 & $-$36.632248 & 8.27$\pm$0.10 & $-$94.39$\pm$0.26 & 1.028$\pm$0.025 & $-$2.418$\pm$0.025 & 1.77$\times 10^5$ & 5.06  \\ 
 NGC 6749 & 286.314056 & 1.899756 & 7.59$\pm$0.21 & $-$58.44$\pm$0.96 &  $-$2.829$\pm$0.028 & $-$6.006$\pm$0.027 & 2.11$\times 10^5$ & 7.09  \\ 
 NGC 6752 & 287.717102 & $-$59.984554 & 4.12$\pm$0.04 & $-$26.01$\pm$0.12 & $-$3.161$\pm$0.022 & $-$4.027$\pm$0.022 & 2.76$\times 10^5$ & 5.27\\
 NGC 6760 & 287.800268 & 1.030466 & 8.41$\pm$0.43 & $-$2.37$\pm$1.27 &  $-$1.107$\pm$0.026 & $-$3.615$\pm$0.026 & 2.69$\times 10^5$ & 5.22 \\ 
 NGC 6809 & 294.998779 & $-$30.964750 & 5.35$\pm$0.05 & 174.70$\pm$0.17 & $-$3.432$\pm$0.024 & $-$9.311$\pm$0.024 & 1.93$\times 10^5$ & 6.95 \\ 
 Pal 11 & 296.310000 & $-$8.007222 & 14.02$\pm$0.51 & $-$67.64$\pm$0.76 & $-$1.766$\pm$0.030 & $-$4.971$\pm$0.028 & 1.19$\times 10^4$ & 7.72 \\ 
 NGC 6838 & 298.443726 & 18.779194 & 4.00$\pm$0.05 & $-$22.72$\pm$0.20 & $-$3.416$\pm$0.025 & $-$2.656$\pm$0.024 & 4.57$\times 10^4$ & 5.23  \\ 
 NGC 7078 & 322.493042 & 12.167001 & 10.71$\pm$0.10 & $-$106.84$\pm$0.30 & $-$0.659$\pm$0.024 & $-$3.803$\pm$0.024 & 6.33$\times 10^5$ & 4.30 \\
  \hline
 \end{tabular}
\end{table*}

\begin{table*}
 \caption{Orbital properties of globular clusters that have
 approximately well defined perigalactic and apogalactic mean 
 time-variations $<\!\dot{r}_{\rm min}\!>$, $<\!\dot{r}_{\rm max}\!>$,
 here determined over 5 Gyr. In Liller 1 only the first Gyr is
 considered. The mean time-variations of energy and angular momentum,  
 $<\!\dot{E}\!>$, $<\!\dot{L}_z\!>$, are determined over 1 Gyr.
 Units of $E$ and $L_z$ are UE=10$^5$ km$^2$ s$^{-2}$, 
 UL=10 kpc km s$^{-1}$. The clusters are listed in order of increasing
 right ascension.}
 \label{prop1}
 \begin{tabular}{cccrrcrrrc}
  \hline
 Cluster & $r_{\rm min_0}$ & $r_{\rm max_0}$ & $<\!\dot{r}_{\rm min}\!>$ & $<\!\dot{r}_{\rm max}\!>$ & $E_0$ & $<\!\dot{E}\!>$ & $L_{z_0}$ & $<\!\dot{L}_z\!>$ & Progenitor  \\
         & (kpc) & (kpc) & (pc Gyr$^{-1}$) & (pc Gyr$^{-1}$) & (UE) & ($10^{-3}$UE Gyr$^{-1}$) &  (UL) & (UL Gyr$^{-1}$) &  \\
  \hline
   NGC 5139  & 1.87 & 6.72 & $-30\pm16$ & $-76\pm15$ & $-$1.65 & $-6.27\pm0.07$ & 52.2 & $-0.955\pm0.003$ & G-E/Seq \\
   NGC 6266  & 0.89 & 2.29 & $-26\pm2$ & $-62\pm7$ & $-$2.16 & $-12.32\pm1.6$ & $-$23.8 & 0.387$\pm0.003$ & M-B \\
   Terzan 2  & 0.17 & 0.81 & $-8\pm4$ & $-36\pm2$ & $-$2.79 & $-29.76\pm9.5$ & 6.7 & $-0.344\pm0.08$ & M-B \\
   Terzan 4  & 0.16 & 0.84 & $-4\pm5$ & $-59\pm7$ & $-$2.77 & $-45.55\pm21$ & $-$4.9 & 0.149$\pm0.05$ & M-B \\
   Liller 1  & 0.094 & 0.81 & $-69\pm8$ & $-359\pm60$ & $-$2.84 & $-418.60\pm130$ & 4.2 & $-3.406\pm1.7$ &  \\
   Terzan 1  & 0.49 & 2.64 & 23$\pm6$ & $-66\pm12$ & $-$2.19 & $-7.66\pm1.56$ & $-$21.7 & $-0.756\pm0.17$ & M-B \\
   Terzan 5  & 0.19 & 1.71 & $-38\pm3$ & $-259\pm64$ & $-$2.40 & $-78.11\pm37$ & $-$8.1 & 0.107$\pm0.25$ & M-B \\
   NGC 6440  & 0.31 & 1.27 & $-28\pm10$ & $-142\pm13$ & $-$2.51 & $-60.67\pm11$ & 3.8 & $-0.436\pm0.12$ & M-B \\
   NGC 6441  & 1.68 & 4.84 & 24$\pm3$ & $-91\pm7$ & $-$1.83 & $-6.46\pm0.1$ & $-$50.6 & $-0.195\pm0.015$ & L-E \\
  Djorg 2   & 0.51 & 0.80 & $-8\pm2$ & $-13\pm5$ & $-$2.62 & $-10.03\pm7.7$ & 14.2 & $-0.209\pm0.08$ & M-B \\
  2MASS-GC01 & 3.79 & 5.16 & 2$\pm5$ & $-44\pm4$ & $-$1.72 & $-2.81\pm0.1$ & $-$99.1 & 0.381$\pm0.12$ &  \\
  NGC 6553   & 2.80 & 3.70 & $-45\pm2$ & $-128\pm5$ & $-$1.89 & $-13.45\pm0.2$ & $-$69.1 & 1.687$\pm0.1$ & M-B \\
  IC 1276   & 3.66 & 7.22 & 8$\pm2$ & $-19\pm2$ & $-$1.59 & $-0.70\pm0.02$ & $-$111.2 & $-0.039\pm0.02$ & M-D \\
  NGC 6749   & 1.38 & 4.95 & 55$\pm3$ & $-77\pm6$ & $-$1.87 & $-3.92\pm0.1$ & $-$50.2 & $-0.974\pm0.04$ & M-D \\
  NGC 6760   & 1.86 & 5.78 & 24$\pm2$ & $-48\pm3$ & $-$1.76 & $-2.44\pm0.22$ & $-$63.4 & $-0.362\pm0.01$ & M-D \\
  NGC 6838   & 4.86 & 7.08 & 3$\pm2$ & $-17\pm2$ & $-$1.54 & $-0.70\pm0.01$ & $-$132.8 & 0.098$\pm0.01$ & M-D \\
  \hline
 \end{tabular}
\end{table*}

\begin{table*}
 \caption{Orbital properties of globular clusters not included in
 Table~\ref{prop1}. The mean time-variations $<\!\dot{E}\!>$, $<\!\dot{L}_z\!>$
 are determined over the first Gyr. Units are those given in
 Table~\ref{prop1}.}
 \label{prop2}
 \begin{tabular}{c@{\hspace{0.15cm}}c@{\hspace{0.15cm}}c@{\hspace{0.15cm}}c@{\hspace{0.15cm}}c@{\hspace{0.17cm}}r@{\hspace{0.17cm}}r@{\hspace{0.17cm}}c@{\hspace{0.17cm}}c@{\hspace{0.15cm}}c@{\hspace{0.15cm}}c@{\hspace{0.15cm}}c@{\hspace{0.15cm}}c@{\hspace{0.17cm}}r@{\hspace{0.17cm}}r@{\hspace{0.17cm}}c}
  \hline
 Cluster & $r_{\rm min_0}$ & $r_{\rm max_0}$ & $E_0$ & $<\!\dot{E}\!>$ & $L_{z_0}$ & $<\!\dot{L}_z\!>$ & Progenitor & Cluster & $r_{\rm min_0}$ & $r_{\rm max_0}$ & $E_0$ & $<\!\dot{E}\!>$ & $L_{z_0}$ & $<\!\dot{L}_z\!>$ & Progenitor  \\
  \hline
 NGC 104  & 5.91 & 7.51 & $-$1.46 & $-$0.554 & $-$124.1 & 0.117 & M-D & NGC 6401  & 0.38 & 1.33 & $-$2.46 & $-$12.011 & 1.9 & $-$0.060 & K \\
   NGC 4372  & 2.91 & 7.22 & $-$1.60 & $-$0.501 & $-$90.4 & $-$0.008 &
 M-D & NGC 6397  & 2.26 & 6.28 & $-$1.66 & $-$0.164 & $-$65.1 & 0.004 &
 M-D \\
  BH 140  & 1.97 & 10.19 & $-$1.49 & $-$0.349 & $-$81.1 & $-$0.045 & &
 Pal 6  & 0.60 & 1.80 & $-$2.25 & $-$3.322 & 2.1 & $-$0.043 & M-B \\
   NGC 4833  & 0.51 & 8.02 & $-$1.63 & $-$0.821 & $-$23.6 & $-$0.033 &
 G-E & Djorg 1  & 1.18 & 10.28 & $-$1.50 & $-$0.320 & $-$53.7 &
 $-$0.038 & G-E \\
   NGC 5927  & 4.22 & 5.68 & $-$1.64 & $-$2.333 & $-$107.7 & 0.372 &
 M-D & Terzan 6  & 0.19 & 1.08 & $-$2.63 & $-$18.257 & 8.5 & $-$0.293 &
 M-B \\
   NGC 5946  & 0.19 & 5.32 & $-$1.82 & $-$0.731 & $-$4.6 & $-$0.016 &
 K & NGC 6453  & 0.67 & 2.14 & $-$2.12 & $-$5.100 & $-$5.8 & $-$0.126 &
 K \\
   NGC 5986  & 1.37 & 5.05 & $-$1.80 & $-$2.518 & $-$13.8 & $-$0.062 &
 K & UKS 1  & 0.18 & 8.09 & $-$1.65 & $-$0.583 & $-$11.5 & $-$0.035 & \\
   FSR 1716  & 2.41 & 5.02 & $-$1.78 & $-$0.245 & $-$65.6 & 0.004 & M-D & VVV CL001  & 0.58 & 1.56 & $-$2.36 & $-$5.016 & 21.1 & $-$0.224 &
 G-E/Seq \\
   Lynga 7  & 1.72 & 4.54 & $-$1.86 & $-$0.474 & $-$51.3 & $-$0.018 &
 M-D & NGC 6496  & 2.45 & 5.29 & $-$1.72 & $-$0.118 & $-$61.9 & 0.004 &
 M-D \\
   NGC 6093  & 1.23 & 4.48 & $-$1.85 & $-$4.088 & $-$7.1 & $-$0.137 &
 K & Terzan 9  & 0.23 & 2.65 & $-$2.20 & $-$3.774 & $-$12.2 & $-$0.004 &
 M-B \\
   NGC 6121  & 0.46 & 6.45 & $-$1.75 & $-$0.692 & $-$23.8 & $-$0.064 &
 L-E & NGC 6517 & 0.63 & 3.29 & $-$2.04 & $-$3.637 & $-$12.3 & $-$0.070 & K \\
   NGC 6144  & 2.08 & 3.30 & $-$1.88 & $-$0.197& 16.6 & $-$0.012 & K &
 Terzan 10 & 0.80 & 5.71 & $-$1.74 & $-$1.069 & $-$20.6 & 0.005 & G-E \\
   NGC 6139  & 0.92 & 3.69 & $-$1.91 & $-$1.354 & $-$26.0 & 0.009 & K & NGC 6522  & 0.33 & 1.15 & $-$2.49 & $-$21.569 & $-$8.2 & 0.100 & M-B \\
  Terzan 3  & 2.26 & 3.04 & $-$1.91 & $-$0.114 & $-$45.2 & 0.011 & M-D & NGC 6535 & 0.98 & 4.58 & $-$1.88 & $-$0.093 & 30.3 & $-$0.009 & Seq \\
   NGC 6171  & 1.52 & 3.94 & $-$1.90 & $-$0.648 & $-$26.6 & $-$0.053 &
 M-B & NGC 6528   & 0.38 & 0.73 & $-$2.70 & $-$10.157 & $-$4.1 & 0.032 &
 M-B \\
 ESO 452-SC11 & 0.26 & 2.81 & $-$2.16 & $-$0.321 & $-$1.8 & $-$0.010 & & NGC 6539 & 1.77 & 3.40 & $-$1.87 & $-$0.468 & $-$33.0 & 0.025 & M-B \\
 NGC 6218  & 2.49 & 4.79 & $-$1.76 & $-$0.207 & $-$46.9 & 0.009 & M-D &
 NGC 6540 & 0.86 & 2.30 & $-$2.18 & $-$0.859 & $-$28.9 & 0.024 & M-B \\
 FSR 1735  & 1.09 & 3.83 & $-$1.93 & $-$0.426 & $-$33.2 & 0.008 & K &
 NGC 6544 & 0.43 & 5.60 & $-$1.82 & $-$0.959 & $-$17.1 & $-$0.046 & K \\
 NGC 6235  & 3.10 & 8.56 & $-$1.48 & $-$0.085 & $-$76.7 & 0.010 & G-E &
 NGC 6541  & 1.51 & 3.68 & $-$1.89 & $-$0.840 & $-$32.8 & 0.028 & K \\
 NGC 6254  & 1.59 & 4.60 & $-$1.79 & $-$0.412 & $-$42.0 & 0.016 & K &
 2MASS-GC02 & 0.80 & 6.28 & $-$1.74 & $-$0.068 & $-$21.2 & $-$0.001 & \\
 NGC 6256  & 1.31 & 2.11 & $-$2.14 & $-$2.540 & $-$34.3 & 0.169 & K &
 NGC 6558  & 0.51 & 1.26 & $-$2.43 & $-$1.836 & $-$6.2 & 0.006 & M-B \\
 NGC 6273  & 0.81 & 3.29 & $-$1.93 & $-$3.416 & 7.3 & $-$0.100 & K & 
Terzan 12 & 1.62 & 3.69 & $-$1.94 & $-$0.728 & $-$47.0 & 0.005 & M-D \\
  NGC 6284 & 0.23 & 6.58 & $-$1.66 & $-$0.713 & 6.6 & $-$0.013 & G-E &   NGC 6569 & 1.76 & 2.83 & $-$2.00 & $-$1.583 & $-$40.1 & 0.116 & M-B \\
 NGC 6287  & 0.23 & 4.48 & $-$1.85 & $-$0.978 & $-$2.9 & $-$0.014 & K &  BH 261  & 1.69 & 2.97 & $-$1.99 & $-$0.142 & $-$39.9 & 0.010 & M-B \\
 NGC 6293  & 0.20 & 2.65 & $-$2.09 & $-$2.881 & 6.7 & $-$0.047 & M-B &
 NGC 6624   & 0.19 & 1.28 & $-$2.46 & $-$15.451 & $-$2.5 & $-$0.057 &
 M-B \\
 NGC 6304  & 1.45 & 2.67 & $-$2.06 & $-$1.482 & $-$38.1 & 0.094 & M-B &  NGC 6626 & 0.33 & 3.11 & $-$2.09 & $-$5.939 & $-$15.3 & $-$0.036 &
 M-B \\
 NGC 6316  & 1.15 & 3.91 & $-$1.94 & $-$1.995 & $-$26.5 & 0.047 & M-B &  NGC 6638   & 0.58 & 2.33 & $-$2.19 & $-$5.011 & $-$2.5 & $-$0.043 &
 M-B \\
 NGC 6325  & 0.94 & 1.31 & $-$2.28 & $-$0.973 & 12.5 & $-$0.027 & M-B &  NGC 6637   & 0.59 & 1.98 & $-$2.23 & $-$5.298 & $-$6.0 & $-$0.026 &
 M-B \\
 NGC 6333  & 0.67 & 6.81 & $-$1.66 & $-$1.079 & $-$25.7 & $-$0.026 &
 K & NGC 6642   & 0.35 & 1.84 & $-$2.32 & $-$2.561 & 1.7 & $-$0.040 & 
 M-B \\
 NGC 6342  & 0.61 & 1.69 & $-$2.24 & $-$0.873 & $-$9.5 & 0.006 & M-B &
 NGC 6652   & 0.26 & 3.34 & $-$2.04 & $-$1.106 & $-$2.6 & $-$0.011 &
 M-B \\
 NGC 6356  & 3.71 & 8.98 & $-$1.46 & $-$0.475 & $-$93.3 & 0.055 & M-D &  NGC 6656  & 3.10 & 9.79 & $-$1.46 & $-$0.698 & $-$93.7 & 0.060 & M-D \\
 NGC 6355  & 0.76 & 1.25 & $-$2.31 & $-$1.879 & 3.9 & $-$0.025 & M-B &
 Pal 8  & 1.20 & 4.19 & $-$1.91 & $-$0.337 & $-$34.8 & 0.008 & M-D \\
 NGC 6352  & 3.12 & 4.12 & $-$1.81 & $-$0.681 & $-$75.2 & 0.091 & M-D &  NGC 6681 & 1.82 & 5.15 & $-$1.76 & $-$1.141 & 0.2 & $-$0.033 & K \\
 NGC 6366  & 2.07 & 5.79 & $-$1.72 & $-$0.127 & $-$64.4 & $-$0.003 &
 M-D & NGC 6712   & 0.29 & 4.74 & $-$1.85 & $-$1.375 & 1.6 & $-$0.057 & K \\
 HP 1  & 0.82 & 1.60 & $-$2.27 & $-$2.716 & 0.4 & $-$0.018 & M-B &
 NGC 6717  & 0.86 & 2.53 & $-$2.13 & $-$0.634 & $-$19.5 & 0.005 & M-B \\
 NGC 6362  & 3.13 & 5.27 & $-$1.69 & $-$0.154 & $-$56.5 & 0.012 & M-D &  NGC 6723 & 2.56 & 2.62 & $-$1.91 & $-$0.448 & $-$4.3 & $-$0.006 & M-B\\
 NGC 6380  & 0.20 & 2.32 & $-$2.26 & $-$13.088 & 5.8 & $-$0.283 & M-B &  NGC 6752  & 3.66 & 5.49 & $-$1.67 & $-$0.501 & $-$83.3 & 0.055 & M-D \\
 Ton 2  & 1.74 & 2.85 & $-$1.98 & $-$0.384 & $-$39.8 & 0.022 & K &
 NGC 6809  & 1.57 & 5.66 & $-$1.70 & $-$0.337 & $-$23.0 & 0.003 & K \\
 NGC 6388  & 1.11 & 4.08 & $-$1.93 & $-$5.750 & 32.5 & $-$0.543 & M-B &  Pal 11  & 4.94 & 8.66 & $-$1.44 & $-$0.008 & $-$121.7 & 0.001 & M-D \\
 NGC 6402 & 0.49 & 4.68 & $-$1.91 & $-$5.233 & $-$14.8 & $-$0.082 & K &
 NGC 7078 & 4.12 & 10.74 & $-$1.38 & $-$0.467 & $-$110.8 & 0.060 &
 M-D \\
  \hline
 \end{tabular}
\end{table*}

\begin{figure}
\includegraphics[width=\columnwidth]{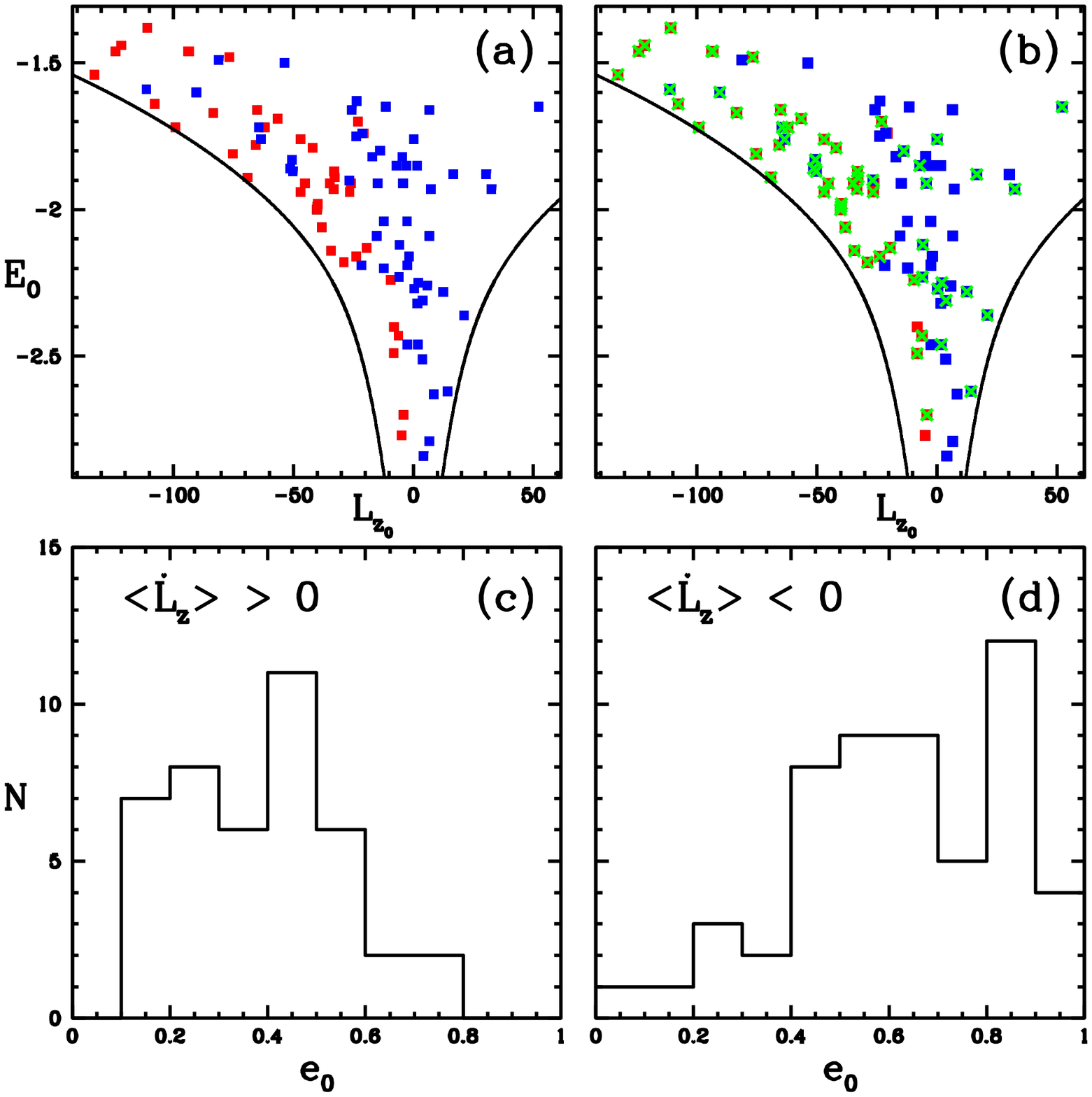}
\caption{Distributions of clusters in the plane ($L_{z_0}$,$E_0$)
and with respect to initial orbital eccentricity. 
In panel (a) the points with red colour correspond to clusters with
positive $<\!\dot{L}_z\!>$, and those in blue colour, to clusters
with negative $<\!\dot{L}_z\!>$.
All the clusters with retrograde motion, i.e. $L_{z_0}$ > 0, have
negative $<\!\dot{L}_z\!>$. Clusters with prograde motion, 
i.e. $L_{z_0}$ < 0, can have $<\!\dot{L}_z\!>$ positive or negative.
The black curves represent prograde and retrograde circular orbits
in the axisymmetric Galactic potential. In panel (b), the points
marked with a green cross correspond to clusters that have an initial
eccentricity, $e_0$, smaller than 0.6. Panels (c) and (d)
show the 
details of the distributions in eccentricity for clusters
with $<\!\dot{L}_z\!>$ positive and negative, i.e. red and blue points
in panel (a).}
\label{fig6}
\end{figure}

With data from Tables~\ref{prop1} and \ref{prop2}, in Fig.~\ref{fig6}
we show how the clusters are distributed in the plane 
($L_{z_0}$,$E_0$), their positive or negative mean time-variation
$<\!\dot{L}_z\!>$ over the first Gyr, and their
distribution in initial orbital eccentricities. In panel (a) the points
with red colour correspond to clusters with positive $<\!\dot{L}_z\!>$,
and those in blue colour, to clusters with negative $<\!\dot{L}_z\!>$.
All the clusters with retrograde motion, i.e. $L_{z_0}$ > 0, have
$<\!\dot{L}_z\!>$ negative, which means that their magnitude of
angular momentum $L_z$ is decreasing. Clusters with prograde motion,
i.e. $L_{z_0}$ < 0, can have $<\!\dot{L}_z\!>$ positive or negative.
Those with positive $<\!\dot{L}_z\!>$ have a decreasing magnitude of
$L_z$, and those with negative $<\!\dot{L}_z\!>$ have an increasing
magnitude of $L_z$. The black curves shown in this panel represent
prograde and retrograde circular orbits in the axisymmetric Galactic
potential. In panel (b), we show again all the points in panel (a),
now marked with a green cross the points corresponding to clusters
that have an initial eccentricity, $e_0$, less than 0.6. We define
$e_0=(r_{\rm max_0}-r_{\rm min_0})/(r_{\rm max_0}+r_{\rm min_0})$. 
Practically all the clusters with positive $<\!\dot{L}_z\!>$, i.e.
the red points in panel (a), have $e_0$ < 0.6. This approximate
value depends on the given definition of $e_0$, which relates 
perigalactic and apogalactic distances around the current time.
Another definition of $e_0$ could be
$e_0=(R_{\rm max_0}-R_{\rm min_0})/(R_{\rm max_0}+R_{\rm min_0})$, with
$R$ distance in cylindrical coordinates. Panels (c) and (d)
show the details of the distributions in eccentricity for clusters
with $<\!\dot{L}_z\!>$ positive and negative, i.e. red and blue points
in panel (a).

\begin{figure}
\includegraphics[width=\columnwidth]{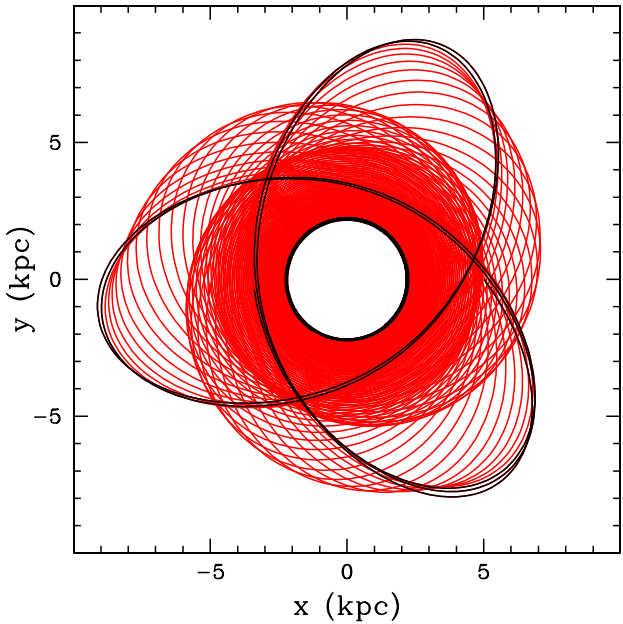}
\caption{Orbit on the Galactic plane of a model cluster with a mass
of $5\times 10^5 M_{\sun}$ and $r_h$=10 pc.
The orbit is initially extended and it becomes nearly circular
at around 15 Gyr as shown in black.} 
\label{fig7}
\end{figure}

\begin{figure}
\includegraphics[width=\columnwidth]{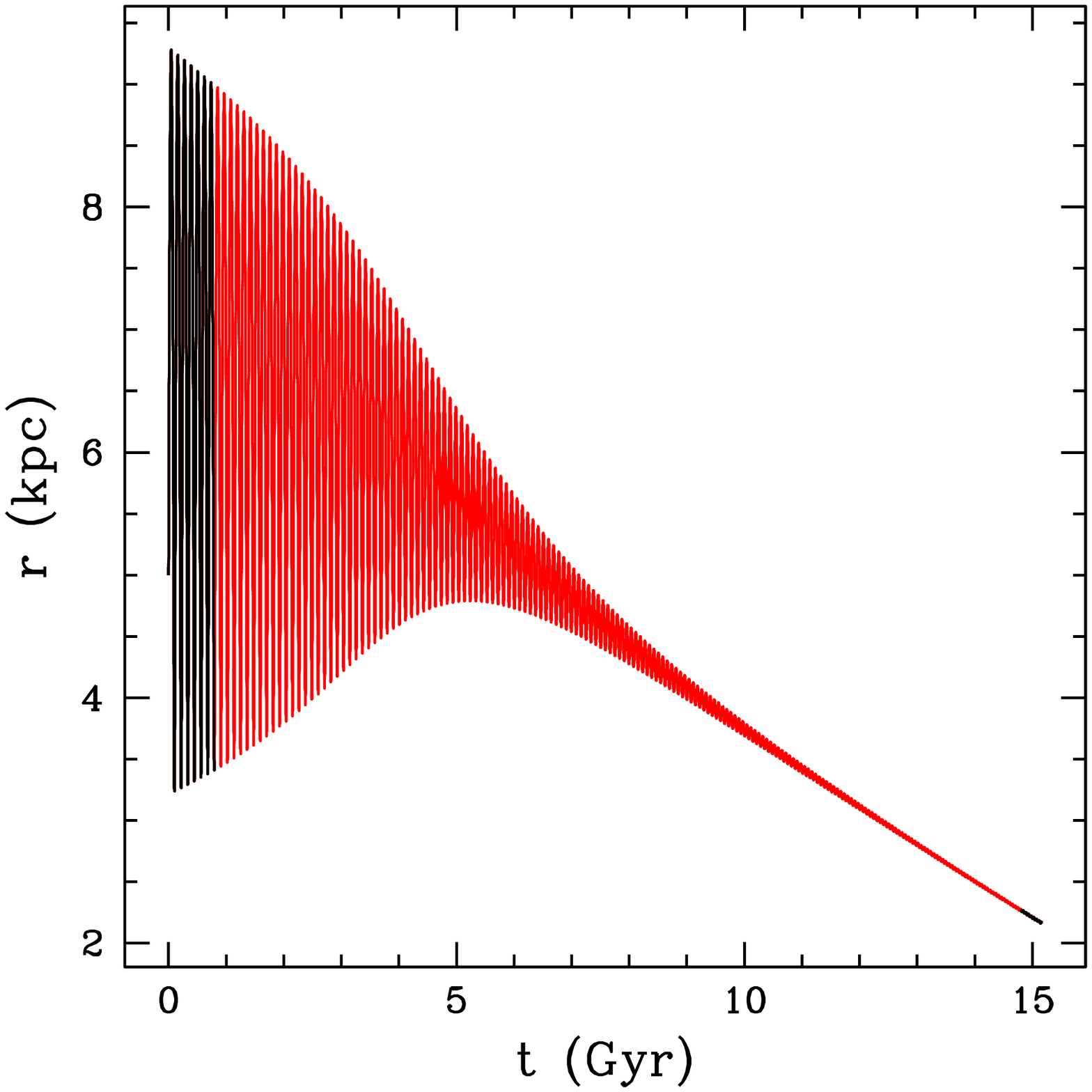}
\caption{The variation of $r$ with time in the orbit shown in
Fig.~\ref{fig7}. The initial and final integration times are
shown in black.} 
\label{fig8}
\end{figure}

All the retrograde clusters have a decreasing $|L_z|$, thus they tend
to increase their eccentricity; as they outnumber those
prograde clusters with negative $<\!\dot{L}_z\!>$, this reflects in the
distribution towards high eccentricity shown in panel (d) of
Fig.~\ref{fig6}. In prograde clusters with decreasing $|L_z|$, i.e.
red points in panel (a) of Fig.~\ref{fig6}, the distribution in
eccentricity shown in panel (c) of Fig.~\ref{fig6} is higher towards
low values. In his computations, \citet{1979A&A....71..245K} found that
dynamical friction acting on a model cluster with low z-amplitude
motion tends to circularize its motion faster than in a cluster with
high z-amplitude, and after that, the cluster has a spiral motion
towards the Galactic Centre, where it may be destroyed. This can
explain in part the lack of clusters with nearly circular orbits in
panel (c). Liller 1, Terzan 4, Terzan 5, and NGC 6440 appear
to be in this phase of fast circularization of their orbits and
possible later destruction. Panel (d) in Fig.~\ref{fig6} shows
that there is only one cluster with $e_0$ < 0.1, this is cluster
NGC 6723 with prograde motion and a decreasing low-magnitude
$|L_{z_0}|$ (see Table~\ref{prop2}). This cluster is resisting to be
dragged towards the Galactic Centre due to its high-amplitude z-motion
reaching $|z|_{\rm max}$ $\approx$ 3 kpc. Another cause for the lack
of nearly circular orbits is the long time needed to attain the
circularization in clusters that cover a wide interval in
Galactocentric distances $r$. For instance, this is the case
in NGC 6441, 2MASS-GC01,
NGC 6749, as estimated from the variations of $r_{\rm min}$,
$r_{\rm max}$ in Figs.~\ref{fig3} and \ref{fig5}. To illustrate
this behaviour, Figs.~\ref{fig7}
and \ref{fig8} show as an example the orbit of a model cluster
with a mass of $5\times 10^5 M_{\sun}$ and $r_h$=10 pc, moving on the
Galactic plane. At initial times the $r$-distances lie within
$\approx$ 3--9 kpc, shown in black colour in both figures, and after
15 Gyr the cluster has a nearly circular orbit with a radius of
$\approx$ 2 kpc, shown also with black colour in both figures.

\subsection{Local variations}
\label{varloc}

\begin{figure}
\includegraphics[width=\columnwidth]{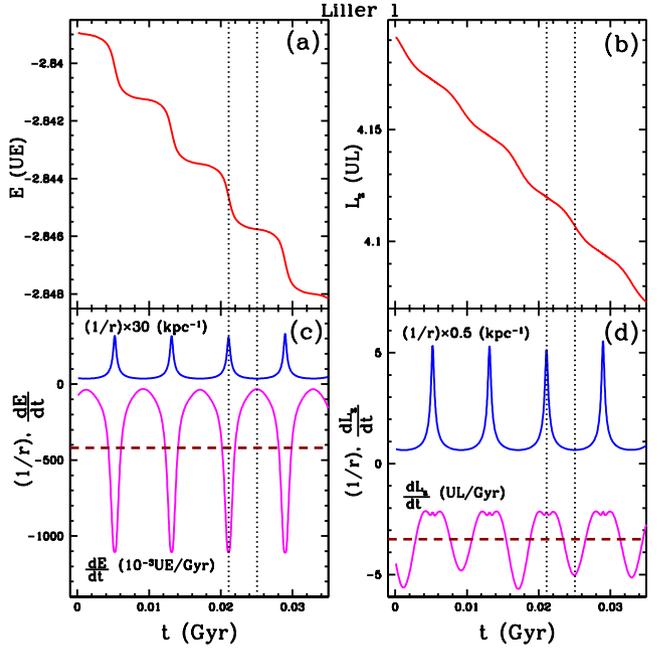}
\caption{Local variations of $E$, $L_z$, $\frac{{\rm d}E}{{\rm d}t}$,
$\frac{{\rm d}L_z}{{\rm d}t}$ in Liller 1
 close to the beginning of its
orbital computation. The units are those given in Table~\ref{prop1}.
Panels (a) and (b) show the smooth variations $E$ vs t and
$L_z$ vs t. Panels (c) and (d)
give $\frac{{\rm d}E}{{\rm d}t}$ vs t, $\frac{{\rm d}L_z}{{\rm d}t}$
 vs t, (magenta colour),
along with the function $\frac{1}{r}$ (blue colour), multiplied by a
convenient factor, to highlight the successive $r_{\rm min}$,
$r_{\rm max}$ positions. The vertical dotted lines extended to panels
(a), (b) localise two positions, one with $r_{\rm min}$ and the other 
with $r_{\rm max}$. Panel (c) shows that $|\frac{{\rm d}E}{{\rm d}t}|_{\rm max}$
occurs at perigalactic distance $r_{\rm min}$.
The minimum value of $|\frac{{\rm d}E}{{\rm d}t}|$ is obtained at
 $r_{\rm max}$. 
The horizontal dashed line shows the mean time-variation
$<\dot{E}>$ for this cluster listed in Table~\ref{prop1}.
Panel (d) shows that $|\frac{{\rm d}L_z}{{\rm d}t}|_{\rm max}$ occurs at
apogalactic distance $r_{\rm max}$, and
$|\frac{{\rm d}L_z}{{\rm d}t}|_{\rm min}$ at $r_{\rm min}$.
The horizontal dashed line gives
the mean time-variation $<\!\dot{L}_z\!>$ listed in Table~\ref{prop1}.}
\label{fig9}
\end{figure}

\begin{figure}
\includegraphics[width=\columnwidth]{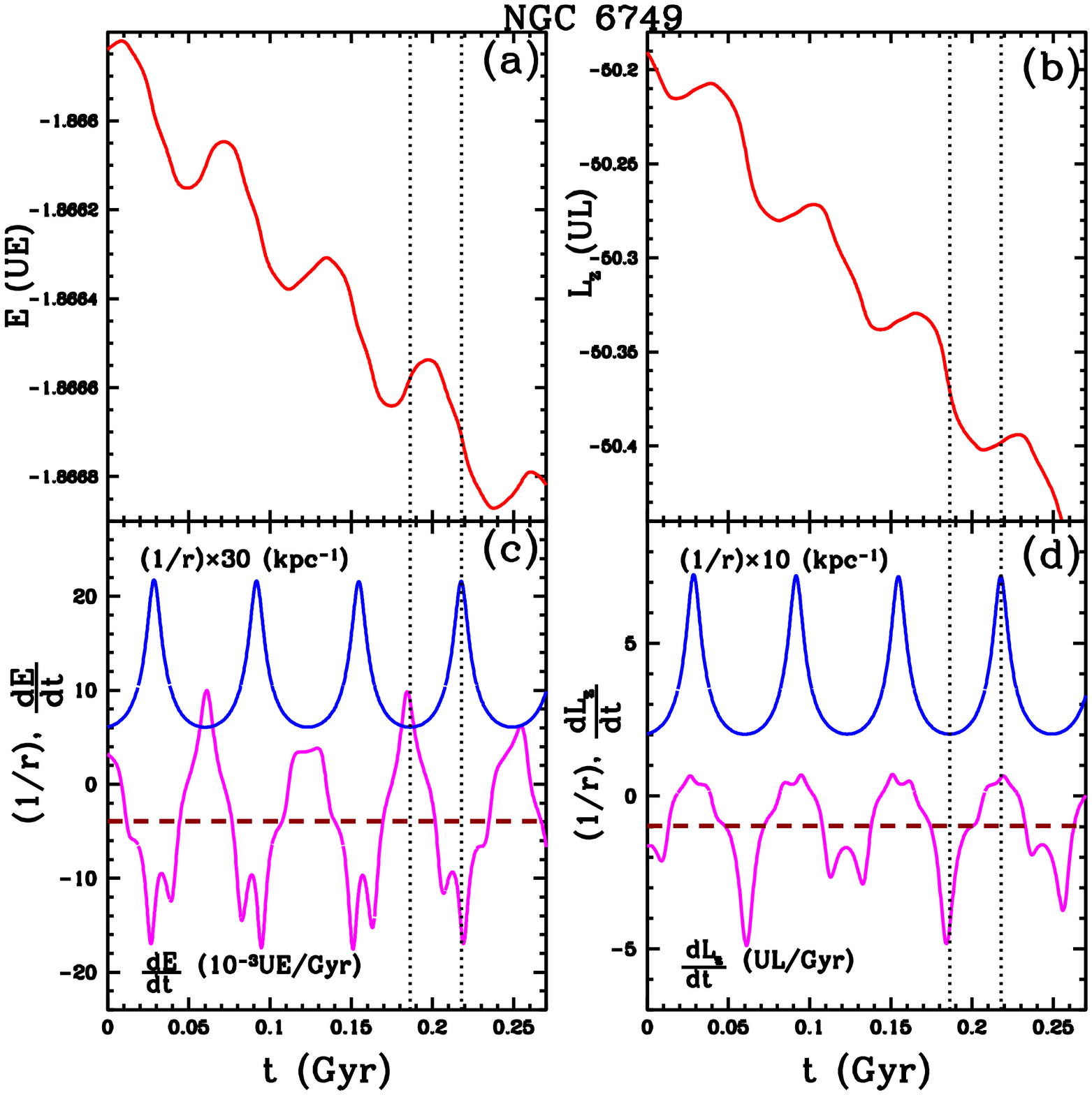}
\caption{As in Fig.~\ref{fig9}, here for NGC 6749.
In this cluster, $E$ and $L_z$ present maxima and minima.
Here $\frac{{\rm d}E}{{\rm d}t}$ can be positive around
 $r_{\rm max}$ with a
value comparable with its maximum magnitude reached at $r_{\rm min}$, 
and $\frac{{\rm d}L_z}{{\rm d}t}$ can change sign at this perigalactic
 distance.}
\label{fig10}
\end{figure}

In Section~\ref{varprom} we computed by means of linear least-square
fits the mean variations $<\dot{E}>$, $<\!\dot{L}_z\!>$ over the first Gyr.
Figs.~\ref{fig1}--\ref{fig5} show smooth variations of $E$ and $L_z$.
However, locally, around a given time, $\frac{{\rm d}E}{{\rm d}t}$ and
$\frac{{\rm d}L_z}{{\rm d}t}$ can be positive and negative, as indicated
for example in the curves $L_z$ vs t of NGC 6171 in
Fig.~\ref{fig1} and NGC 6441 in Fig.~\ref{fig3}. To illustrate the
possible behaviour of $E$, $L_z$, $\frac{{\rm d}E}{{\rm d}t}$, and
$\frac{{\rm d}L_z}{{\rm d}t}$, Figs.~\ref{fig9} and \ref{fig10} show
their dependence on time in Liller 1 and NGC 6749, in a 
short time interval at the beginning of the orbital
computation.
Panels (a) and (b) in both figures show $E$ vs t and $L_z$ vs t.
The details in these curves can not be seen in the corresponding
Figs.~\ref{fig2} and \ref{fig5}, which cover the total 5 Gyr interval.
In Liller 1, $E$ and $L_z$ decline smoothly, but have
maxima and minima in NGC 6749. Panels (c) and (d) in both figures
give $\frac{{\rm d}E}{{\rm d}t}$ vs t, $\frac{{\rm d}L_z}{{\rm d}t}$
 vs t, (magenta colour),
along with the function $\frac{1}{r}$ (blue colour), multiplied by a
convenient factor, to highlight the successive $r_{\rm min}$,
$r_{\rm max}$ positions. The vertical dotted lines extended to panels
(a),(b) localise two positions, one with $r_{\rm min}$ and the other
with $r_{\rm max}$. In both clusters, panels (c) show that
$|\frac{{\rm d}E}{{\rm d}t}|_{\rm max}$ occurs at perigalactic distance
 $r_{\rm min}$.
The minimum value of $|\frac{{\rm d}E}{{\rm d}t}|$ is obtained at
 $r_{\rm max}$ in
Liller 1, but in NGC 6749 $\frac{{\rm d}E}{{\rm d}t}$ can be positive
 around
this apogalactic distance, with a value comparable with the magnitude
of the variation reached at $r_{\rm min}$. The horizontal 
discontinuous lines in panels (c) show the mean time-variations  
$<\dot{E}>$ for these two clusters listed in Table~\ref{prop1}.
In both clusters, panels (d) show that $|\frac{{\rm d}L_z}{{\rm d}t}|_{\rm max}$
occurs at apogalactic distance $r_{\rm max}$, and
$|\frac{{\rm d}L_z}{{\rm d}t}|_{\rm min}$ at $r_{\rm min}$; in NGC 6749
$\frac{{\rm d}L_z}{{\rm d}t}$ can change sign at this perigalactic
 distance.
Also in these panels (d), the horizontal discontinuous lines give
the mean time-variations $<\!\dot{L}_z\!>$ listed in Table~\ref{prop1}.

The results shown in Figs.~\ref{fig9} and \ref{fig10} can be explained
as follows: Liller 1 has a retrograde motion (see
Table~\ref{prop1}), thus it is slowed down along its orbit due to the
prograde motion of the background, and $\frac{{\rm d}E}{{\rm d}t}$,
$\frac{{\rm d}L_z}{{\rm d}t}$ are always negative.
 $|\frac{{\rm d}E}{{\rm d}t}|$ is maximum
at $r_{\rm min}$ due to the increase in density towards the Galactic
Centre. $|\frac{{\rm d}L_z}{{\rm d}t}|$ is maximum at $r_{\rm max}$
 because there
the cluster has a slow motion and the background is moving fast in
the opposite direction. At $r_{\rm min}$ the cluster is moving faster,
but the background coming in the opposite direction has a slow motion,
resulting in less deceleration. In NGC 6749, with a prograde motion
(see Table~\ref{prop1}), at $r_{\rm min}$ the cluster is moving faster
than the background moving in the same direction, thus it is slowed
down, it has less rotation and then $\frac{{\rm d}E}{{\rm d}t}$ is
 negative
and $\frac{{\rm d}L_z}{{\rm d}t}$ positive. At $r_{\rm max}$, the
 cluster is
moving slower than the background coming in the same direction, and
it is accelerated, thus $\frac{{\rm d}E}{{\rm d}t}$ is positive and
 $|L_z|$ increases.

\section{Some implications for the system of globular clusters}
\label{impl}

The globular cluster orbits of Figs.~\ref{fig1}-\ref{fig5} show a wide
range of effects due to dynamical friction, ranging from slight
to moderate and up to extreme, such as for Liller 1, Terzan 4,
Terzan 5, and NGC6440 where the orbits are seen to collapse both
vertically as well as radially. These extreme cases may lead to very
interesting results for analyses of the globular-cluster system and for
Galactic evolution, such as the misclassification of the globulars
between the categories \lq halo', \lq bulge', and \lq thick disc'
\citep{2020MNRAS.491.3251P}, the
resulting biasing of globular-cluster samples, the incorrect
association of the globulars with their parent dwarf galaxies for
accretion events, and the possible formation of nuclear star
clusters.

Globular clusters which had been part of the Galactic halo, with low 
metallicities and early formation in the Galaxy might now, due to 
dynamical friction, appear as members of the Galactic thick disc or
bulge due to their new positions, velocities, and/or distributions in
the Galaxy, as clearly appreciated for Liller 1, Terzan 4, Terzan 5,
and NGC6440.  These might very well be misclassified, and so samples 
of globular clusters for studying these components of the Galaxy: 
thick-disc, bulge, and halo, will be contaminated and biased. 
For example, globulars formed with the ages and metallicities of the 
halo might now be included, due to dynamical friction, in samples for 
the thick disc or bulge, biasing results for age and metallicity 
gradients, as well as for other characteristics, for the bulge, thick
disc, and halo.  The former two components will have some members with
too low metallicities and perhaps too high ages, while the halo sample
will be lacking some members.

\begin{figure}
\includegraphics[width=\columnwidth]{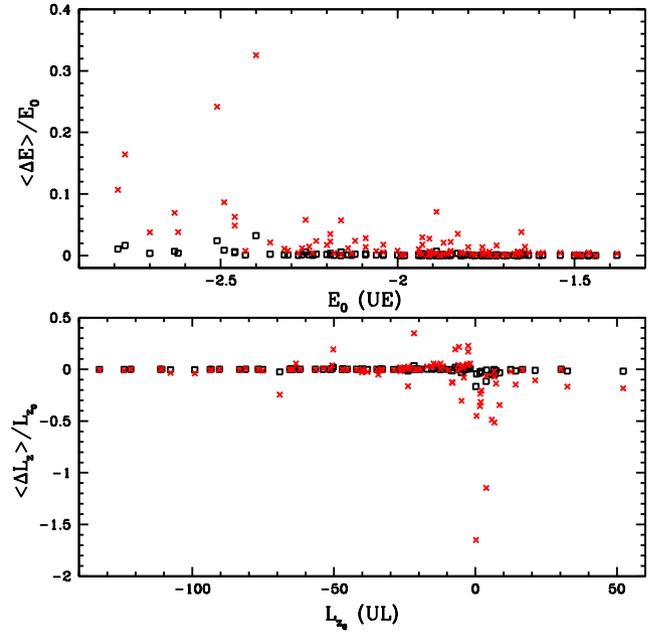}
\caption{Relative variations in energy and angular momentum
of clusters in Tables~\ref{prop1} and \ref{prop2}, except Liller 1. 
Black open squares show the \textit{current} variations, over the first
Gyr of the orbital computations. The red points show the variations
extending the current values over an interval of 10 Gyr.}
\label{fig11}
\end{figure}

For accreted globular clusters, the finer details of the
 hierarchical
galaxy formation could be affected, such as not being able to
separate cleanly their origin. Several studies have associated an
accreted or formed-in-situ origin of some globular clusters 
\citep{2003AJ....125..188B,2009ApJ...702.1058Z,2010MNRAS.404.1203F,
 2010ApJ...721..738Z,2010A&A...511L..10N,2012A&A...538A..21S,
2018ApJ...863L..28M,2018Natur.563...85H,2019MNRAS.488.1235M,
2019A&A...630L...4M,2019A&A...625A...5K,2020MNRAS.493..847F}, but due
to the not considerd effect of dynamical friction, this association
might be uncertain. Regarding this issue, in the literature it has
been usually assumed that in a cluster orbital evolution the orbital
energy and $z$-component of angular momentum are constants, or vary
slightly in time. This approximation has been employed to associate
an accretion event with a possible accreted cluster. With the
results listed in Tables~\ref{prop1} and \ref{prop2}, obtained under
the effect of dynamical friction, we can estimate
if this assumption is convenient. For all the clusters in both tables,
Fig~\ref{fig11} shows the relative variations
$<\Delta E>/E_0$, $<\!\Delta L_z\!>/L_{z_0}$. Liller 1 is
not plotted in this figure since it is off scale, i.e. outside the
vertical limits. Black open squares give the \textit{current}
variations, obtained over the first Gyr of the orbital
computations, as obtained directly from the tables; in energy they are 
approximately smaller than 4 per cent, and in angular momentum the
majority of clusters have variations smaller than 10 per cent.
The red points in the
figure are the corresponding relative variations extending the current
values over a time interval of 10 Gyr. In this extension, in the
majority of clusters the variation in energy is within 10 per cent,
but in angular momentum it can reach about 50 per cent, or more, in
some of them.

The extension to 10 Gyr gives an estimate of expected cluster 
orbital variations in energy and angular momentum over the lifetime of
our Galaxy, applied to possible accreted globular clusters, and those
formed in situ as well. In this extension, the clusters with 
relative variations in energy greater than 10 per cent are: Terzan 2,
Terzan 4, Terzan 5, Liller 1, and NGC 6440. With relative variations
in angular momentum greater than 10 per cent:
NGC 5139, NGC 6093, NGC 6266, NGC 6273, NGC 6380, NGC 6388,
NGC 6401, NGC 6440, NGC 6453, NGC 6522, NGC 6553, NGC 6624, NGC 6638,
NGC 6642, NGC 6681, NGC 6712, NGC 6749, Terzan 1, Terzan 2, Terzan 4,
Terzan 5, Terzan 6, Liller 1, Djorg 2, Pal 6, VVV-CL001, HP 1. 

\begin{figure}
\includegraphics[width=\columnwidth]{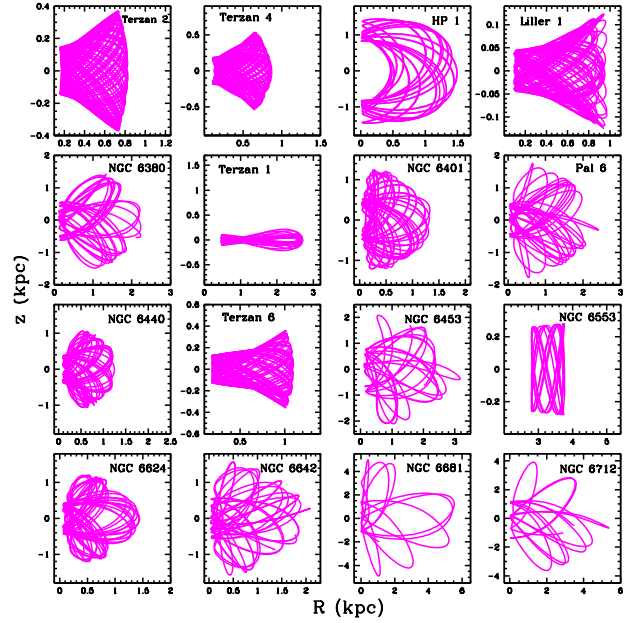}
\caption{Meridional orbits of clusters with variations in
angular momentum greater than 20 per cent, in the extension to 10 Gyr.
The orbits are computed backwards in time up to 10 Gyr, and the figure 
shows the orbits in the first 0.5 Gyr, i.e. near the current time.
The name of the cluster is shown in each panel.}
\label{fig12}
\end{figure}

\begin{figure}
\includegraphics[width=\columnwidth]{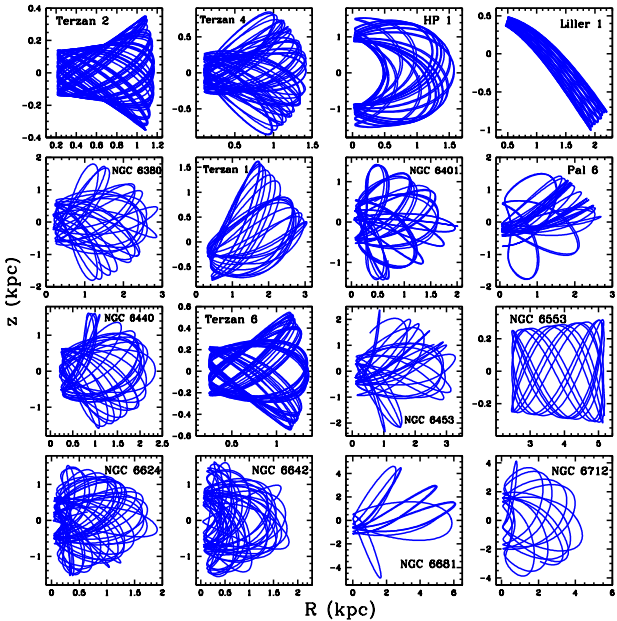}
\caption{Meridional orbits of the clusters in
Fig.~\ref{fig12}, here shown in the time interval from 9.5 Gyr
 to 10 Gyr in the past. Notice the same
scale in the panels as in Fig.~\ref{fig12}, except in Liller 1.}
\label{fig13}
\end{figure}

From these clusters with high variation in angular momentum, 
we separated in particular those with variations greater than 20 per
cent. We computed their orbits backwards in time, i.e. in the
past, up to 10 Gyr, with the effect of the dynamical friction and with the
approximation of constant cluster mass, equal to the current
mass. This underestimates
the dynamical friction effect, as a cluster is more massive in the past.
Also, as in all our computations, the change in time of the
background Galactic potential is not considered.
 In Fig.~\ref{fig12} we show
their meridional orbits in the first 0.5 Gyr of the computations,
i.e. near the current time. Each panel shows the name of the cluster.
Fig.~\ref{fig13} shows the orbits of these same clusters in the last
0.5 Gyr, i.e. from 9.5 Gyr to 10 Gyr in the past; the scale in each panel is the
same as in Fig.~\ref{fig12}, except in Liller 1.
Table~\ref{varEh} lists the relative variations
in energy and $z$-angular momentum in these clusters, during the
total interval of 10 Gyr. In energy the variation in energy is less
than 10 per cent, except in Liller 1. Only Terzan 2, HP 1, Liller 1,
NGC 6380, NGC 6401, NGC 6440, and NGC 6681 present in angular
momentum the estimated variation greater than 20 per cent.
Thus, for the majority of
computed clusters the approximation of nearly invariant energy and 
$z$- angular momentum seems appropriate. 

\begin{table}
\caption{Relative variations in energy and angular momentum for the
clusters in Figs.~\ref{fig12} and \ref{fig13}, during 10 Gyr
backwards in time with the effect of dynamical friction.}
\label{varEh}
\begin{tabular}{ccc}
\hline
 Cluster  & $<\!\Delta E\!>/E_0$ & $<\!\Delta L_z\!>/L_{z_0}$  \\
\hline
 Terzan 2 &  $-$0.040 &  0.225  \\
 Terzan 4 &  $-$0.056 &  0.051  \\
 HP 1     &  $-$0.006 &  0.223  \\
 Liller 1 &  $-$0.150 &  2.126  \\
 NGC 6380 &  $-$0.024 &  0.326  \\
 Terzan 1 &  $-$0.016 &  $-$0.175  \\
 NGC 6401 &  $-$0.024 &  0.317  \\
 Pal 6    &  $-$0.009 &  0.139  \\
 NGC 6440 &  $-$0.067 &  0.718  \\
 Terzan 6 &  $-$0.028 &  0.157  \\
 NGC 6453 &  $-$0.009 &  $-$0.050  \\
 NGC 6553 &  $-$0.029 &  0.070  \\
 NGC 6624 &  $-$0.026 &  $-$0.144  \\
 NGC 6642 &  $-$0.005 &  0.084  \\
 NGC 6681 &  $-$0.003 &  0.712  \\
 NGC 6712 &  $-$0.003 &  0.137  \\
\hline
\end{tabular}
\end{table}

Figs.~\ref{fig12} and \ref{fig13} show that Terzan 2, Terzan 4,
Terzan 6, NGC 6624, NGC 6642, HP 1, which have been classified
as formed in situ in the Galactic bulge 
(see Tables~\ref{prop1} and \ref{prop2}), approximately maintain this
classification over the 10 Gyr computed time. This appears only
approximate in Terzan 1, NGC 6440, NGC 6380, Pal 6, with the same
classification. However, NGC 6553 appears to come from a region
external to the Galactic bulge, i.e. not formed in situ within this
bulge. Liller 1 shows the strongest variation, and it needs a
particular analysis. NGC 6401, NGC 6453, NGC 6681, NGC 6712,
are associated with the $Koala$ dwarf galaxy accretion event
\citep{2020MNRAS.493..847F}, and their orbits do not change
strongly during 10 Gyr, in spite of the strong variation in angular
momentum in NGC 6401 and NGC 6681. The results shown in
Figs.~\ref{fig12} and \ref{fig13} are preliminary, as we have assumed
a constant cluster mass. A study varying the mass of clusters, similar
to that presented by \citet{2019MNRAS.482.5138B}, could improve the
comparison.

Another process where globular clusters extremely affected 
by dynamical friction might be
pertinent for understanding Galactic observations, would be globulars 
infalling toward the inner-most region of the
Galaxy, and in many other galaxies as well, due to their rapidly
collapsing orbits. This process could form nuclear star clusters (NSCs)
 \citep{1975ApJ...196..407T,1993ApJ...415..616C,
2000ApJ...543..620O,2001ApJ...552..572L,2011ApJ...729...35A,
2013MmSAI..84..167C,2012ApJ...750..111A,2014ApJ...784L..44P,
2014ApJ...785...71G,2014CQGra..31x4007S,2015ApJ...806..220A,
2017MNRAS.464.3720T,2020A&ARv..28....4N}
and distributions of clusters around the nuclear galactic regions.
Some massive accreted clusters in the Galaxy are themselves
NSCs of accreted dwarf galaxies \citep{2021MNRAS.500.2514P}.
These infalling clusters might induce new star formation in the
material near the central black hole, showing 
both very young as well as very old stars, as observed.

\section{Conclusions}
\label{concl}

We have made a preliminary analysis of the effect of dynamical friction
on some orbits of globular clusters in our Galaxy, considering an
anisotropic velocity dispersion field approximated with some studies
in the literature. An axisymmetric Galactic model with mass components
disc, bulge, and dark halo is employed in the computations.
We describe a procedure to
compute the dynamical friction acceleration in ellipsoidal, oblate,
and prolate velocity distribution functions with similar density in
velocity space. Orbital properties, such as mean time-variations of
perigalactic and apogalactic distances, energy, and z-component of
angular momentum, are obtained for globular clusters
lying in the Galactic region $R \lesssim$ 10 kpc, $|z| \lesssim$ 5 kpc,
with $R,z$ cylindrical coordinates. These include clusters in
prograde and retrograde orbital motion. Several clusters are strongly
affected by dynamical friction, in particular, Liller 1, 
Terzan 4, Terzan 5, NGC 6440, and NGC 6553, lying in the Galactic inner
region.
Improvements to our computations can be made analysing with more
detail the dispersion fields of the Galactic mass components,
specially the thin and thick disc components. Introducing the
properties of the dark halo outside galactocentric distance of
10 kpc will permit the analysis of the orbits of globular clusters not
included in the present study. More extended analyses including these
improvements will be presented in the future, using the detailed
Galactic model GravPot16 (see footnote 1 in Section~\ref{modgal}),
considering the computation of tidal radii, destruction rates, and
other properties of globular clusters related with the effect of
dynamical friction.

We have pointed out some aspects in the analysis of the system of
Galactic globular clusters which are directly related with the effect
of dynamical friction. These include the possible misclassification 
of the globulars between the categories \lq halo', \lq bulge', and
\lq thick disc', the resulting biasing of globular-cluster samples, the
possible incorrect association of the globulars with their parent
dwarf galaxies for accretion events, and the possible formation of
\lq nuclear star clusters'.

\section*{Acknowledgements}

We thank the referee for comments and suggestions which improved this
paper. JGF-T gratefully acknowledges grants support from Comit\'e
Mixto ESO-Chile 2021.
AP-V acknowledges the DGAPA-PAPIIT grant IG100319.
LC-V acknowledges the support of the Postdoctoral Fellowship of
DGAPA-UNAM, M\'exico, and the Fondo Nacional de Financiamiento para la
Ciencia, la Tecnolog\'ia y la Innovaci\'on `FRANCISCO JOS\'E DE
CALDAS', MINCIENCIAS, and the VIIS for the economic support
of this research. WJS wishes to thank DW Schuster for technical
support: at home computing, the handling of emails, and the editing
of electronic manuscripts.

\section*{Data Availability}

Data available on request.



\bibliographystyle{mnras}




\appendix

\section{Force field of an ellipsoidal shell in velocity space with
linear similar density}
\label{ap1}

As in any continuous function, a density function $f_0$ in velocity
space whose isodensity surfaces are similar ellipsoids, as in the case
of $f_0$({\boldmath $r$},{\boldmath $v$}) in Eq.~\ref{fd0},
can be approximated by a series of connected linear segments, which are
linear functions of an appropriate variable identifying the ellipsoidal
surfaces. Thus, each linear segment represents a shell in velocity
space, and the original density function can be reasonably
approximated by taking a sufficiently large number of these shells.
The total potential ${\Psi}$({\boldmath $r$},{\boldmath $v$}$_{\rm M}$) (see Eq.~\ref{fh0}) and force field
$-{\nabla}$$_{\tiny v_{\rm M}}$${\Psi}$({\boldmath $r$},{\boldmath $v$}$_{\rm M}$) at point {\boldmath $v$}$_{\rm M}$ 
are obtained with the sum of the individual fields of each shell.

We denote the semiaxes of an ellipsoidal surface in velocity space with
the usual notation in ordinary space: $a, b, c$, along the
Cartesian axes $({v}_1,{v}_2,{v}_3)$,
respectively, with $a$>$b$>$c$. The unitary vectors along the axes
are ({\boldmath $e$}$_1$,{\boldmath $e$}$_2$,{\boldmath $e$}$_3$).
Taking the major semi-axis $a$ as the velocity variable which
identifies the similar
ellipsoidal surfaces, the linear density in a given shell with
$a$ $\epsilon [a_1,a_2]$ has the form $f_0(a)=p_0+p_1a$, with
$p_0, p_1$ constants which are obtained in terms of $f_0$ 
evaluated at $a_1, a_2$. With $a=ma_2$, $m$ $\epsilon [m_1,1],
a_1=m_1a_2$, the density in the shell is $f_0[m]=p_0+p_1a_2m$, and its 
potential at any point {\boldmath $v$}=$({v}_1,{v}_2,{v}_3)$
can be obtained with potential theory \citep[e.g.,][]{Kell53}.
The given point {\boldmath $v$} lies on a similar
ellipsoidal surface with major semi-axis denoted by
$a$({\boldmath $v$}) in the following analysis.\\

\noindent (a) {\boldmath $v$} external to the shell, i.e. 
$a$({\boldmath $v$})$\geq a_2$.\\

\noindent In the region 
external to the shell, i.e. $a$({\boldmath $v$})$\geq a_2$, the
potential is (in the following we omit in the potential its 
dependence on {\boldmath $r$})

\begin{equation} {\Psi}(\mbox{\boldmath $v$}) = -2\pi a_2^3\zeta\xi
\int_{m_1}^{1} f_0[m]m^2{\rm d}m \int_{\lambda(m)}^{\infty} \frac{{\rm d}s}{
\sqrt{\varphi(m,s)}},
\label{pot}
\end{equation}

\noindent where the function $\varphi(m,s)$ is given by

\begin{equation} \varphi(m,s)=(m^2a_2^2+s)(m^2\zeta^2a_2^2+s)(
m^2\xi^2a_2^2+s),
\label{fims}
\end{equation}

\noindent the factors $\zeta=b/a$, $\xi=c/a$ are constant ratios in the
similar ellipsoids, and $\lambda(m)$ is the greatest root of
$s$ in the cubic equation resulting from

\begin{equation} \frac{{v}_1^2}{m^2a_2^2+s}+\frac{{v}_2^2}{m^2\zeta^2a_2^2+s}+
\frac{{v}_3^2}{m^2\xi^2a_2^2+s}-1= 0.
\label{eccubs}
\end{equation}

\noindent Introducing $f_0[m]=p_0+p_1a_2m$ in Eq.~(\ref{pot}), with
the change of variable $s=m^2t$ in the inner integral, and integrating
by parts, we obtain

\begin{eqnarray} {\Psi}(\mbox{\boldmath $v$}) &=& {\Psi}_{p_0}
(\mbox{\boldmath $v$})+ {\Psi}_{p_1}(\mbox{\boldmath $v$})=
 -\pi p_0 a_2^3\zeta\xi(I_1-m_1^2I_2+I_3)-  \nonumber \\    
& &  -
 \frac{2}{3}\pi p_1a_2^4\zeta\xi(I_1-m_1^3I_2+I_4),
\label{pott}
\end{eqnarray}

\noindent with

\begin{equation} I_1= \int_{\lambda}^{\infty} \frac{{\rm d}t}{
\sqrt{\varphi(t)}}, ~~~~~ I_2= \int_{u(m_1)}^{\infty} \frac{{\rm d}t}
{\sqrt{\varphi(t)}}, 
\label{I1I2}
\end{equation}

\begin{equation} I_3= \int_{u(m_1)}^{\lambda} \frac{m^2(u){\rm d}u}{
\sqrt{\varphi(u)}}, ~~~~~ I_4= \int_{u(m_1)}^{\lambda}
\frac{m^3(u){\rm d}u}{\sqrt{\varphi(u)}},
\label{I3I4}
\end{equation}

\begin{equation} \varphi(t)=(a_2^2+t)(\zeta^2a_2^2+t)(\xi^2a_2^2+t),
\label{fit}
\end{equation}

\begin{equation} m(u)= \left[ \frac{{v}_1^2}{a_2^2+u}+\frac{{v}_2^2}
{\zeta^2a_2^2+u}+\frac{{v}_3^2}{\xi^2a_2^2+u} \right]^{1/2},
\label{mu}
\end{equation}

\noindent and $u(m_1)=\lambda(m_1)/m_1^2$, $\lambda=\lambda(m=1)$.
Now, with

\begin{align}
\sin \phi_1&=a_2 \left[\frac{1-\xi^2}{a_2^2+u(m_1)}\right]^{1/2}, \label{sf1}\\
\sin \phi_2&=a_2 \left[\frac{1-\xi^2}{a_2^2+\lambda}\right]^{1/2}, \label{sf2}\\
k&=\left[\frac{1-\zeta^2}{1-\xi^2} \right]^{1/2}, \label{k}
\end{align}

\noindent the integrals $I_1$, $I_2$ are

\begin{equation} I_1= \frac{2F(k,\phi_2)}{a_2\sqrt{1-\xi^2}},
~~~ I_2= \frac{2F(k,\phi_1)}{a_2\sqrt{1-\xi^2}},
\label{solI1I2}
\end{equation}

\noindent with $F(k,\phi)$ being the Legendre elliptic integral of the
1st kind.

\noindent Introducing Eq.~(\ref{mu}) in the integrand of $I_3$, this
integral $I_3$ is a combination of the following three integrals,
where we use expressions in (\ref{sf1}), (\ref{sf2}), (\ref{k}),
 with $k'=\sqrt{1-k^2}$,
and $E(k,\phi)$ being the Legendre elliptic integral of the 2nd kind

\begin{eqnarray} I_5 &=& \int_{u(m_1)}^{\lambda} \frac{{\rm d}u}
{(a_2^2+u)\sqrt{\varphi(u)}}= \frac{2}{a_2^3(1-\xi^2)^{3/2}k^2}
\left[F(k,\phi_1)-\right. \nonumber \\
& &  \left. -F(k,\phi_2)-
[E(k,\phi_1)-E(k,\phi_2)] \right],
\label{I5}
\end{eqnarray}

\begin{eqnarray} I_6 &=& \int_{u(m_1)}^{\lambda} \frac{{\rm d}u}
{(\zeta^2a_2^2+u)\sqrt{\varphi(u)}}= \cr
&=&  \frac{2}{a_2^3(1-\xi^2)^{3/2}k^2k'^2}
\left. \bigg\{k'^2[F(k,\phi_2)-F(k,\phi_1)]-\right. \nonumber \\
& & \left. -
k^2 \left[\frac{\sin\phi_1\cos\phi_1}{\sqrt{1-k^2\sin^2\phi_1}}-
\frac{\sin\phi_2\cos\phi_2}{\sqrt{1-k^2\sin^2\phi_2}} \right]+
 \right. \nonumber \\
& & \left. +
E(k,\phi_1)-E(k,\phi_2) \right. \bigg\},
\label{I6}
\end{eqnarray}


\begin{eqnarray} I_7 &=& \int_{u(m_1)}^{\lambda} \frac{{\rm d}u}
{(\xi^2a_2^2+u)\sqrt{\varphi(u)}}= \cr
&=& \frac{2}{a_2^3(1-\xi^2)^{3/2}k'^2}
\left[\sqrt{1-k^2\sin^2\phi_1}\tan\phi_1-\right. \nonumber \\
& &  
-\sqrt{1-k^2\sin^2\phi_2}\tan\phi_2-  \nonumber \\
& &  -[E(k,\phi_1)-E(k,\phi_2)]  \Bigg].
\label{I7}
\end{eqnarray}

\noindent One way to solve the remaining integral $I_4$ is to note
that from expressions (\ref{fit}) and (\ref{mu}),
$m(u)\sqrt{\varphi(u)}$ can be written as

\begin{equation} m(u)\sqrt{\varphi(u)}=(Au^2+Bu+C)^{1/2},  
\label{mrafi}
\end{equation}

\noindent with $A, B, C$ functions of $({v}_1,{v}_2,{v}_3)$.
 Thus, writing the
integrand in $I_4$ as $m^4(u)/(Au^2+Bu+C)^{1/2}$, $I_4$ can be obtained
as the sum of six integrals easily computed with standard tables
\citep[e.g.,][]{GR00}. This completes the computation of the
potential in Eq.~(\ref{pott}).

\noindent The force field is obtained with the negative gradient of the 
potential. For the part ${\Psi}_{p_0}$({\boldmath $v$}) in 
Eq.~(\ref{pott}) we have

\begin{eqnarray} -\frac{\partial {\Psi}_{p_0}}{\partial {v}_1}
 &=&
\pi p_0a_2^3\zeta\xi \left[-\frac{1}{\sqrt{\varphi(\lambda)}}\frac
{\partial \lambda}{\partial {v}_1}+\frac{m_1^2}{\sqrt{\varphi(u(m_1))}}
\frac{\partial u(m_1)}{\partial {v}_1}+\right. \nonumber \\
& &  
+\int_{u(m_1)}^{\lambda} \frac{\partial m^2(u)}{\partial {v}_1} \frac{{\rm d}u}
{\sqrt{\varphi(u)}}+\frac{m^2(\lambda)}{\sqrt{\varphi(\lambda)}}\frac
{\partial \lambda}{\partial {v}_1}-  \nonumber \\
& &  \left. -
\frac{m^2(u(m_1))}{\sqrt{\varphi(u(m_1))}}\frac{\partial u(m_1)}{\partial {v}_1} \right ].
\label{fxfipoo}
\end{eqnarray}

\noindent The sum of the first and fourth terms inside the parenthesis
in this last equation is zero, because from Eqs.~(\ref{eccubs}) and
(\ref{mu}) $\lambda$ is a root of $m^2(u)-1=0$. Also, the sum of the
second and last terms is zero; this sum has the factor

\begin{eqnarray} m_1^2-m^2(u(m_1)) &=& m_1^2 \left [1-
\frac{{v}_1^2}{a_1^2+\lambda(m_1)}-\frac{{v}_2^2}{\zeta^2a_1^2+\lambda(m_1)}-
\right. \nonumber \\
& &  \left. -
\frac{{v}_3^2}{\xi^2a_1^2+\lambda(m_1)} \right],
\label{dif}
\end{eqnarray}

\noindent and with Eq.~(\ref{eccubs}) $\lambda(m_1)$ cancels the
 expression inside the
parenthesis. Thus, with expressions (\ref{mu}) and (\ref{I5})

\begin{equation} -\frac{\partial {\Psi}_{p_0}}{\partial {v}_1}=
2\pi p_0a_2^3\zeta\xi {v}_1I_5. 
\label{fxfipo}
\end{equation}

\noindent Analogously

\begin{equation} -\frac{\partial {\Psi}_{p_0}}{\partial {v}_2}=
2\pi p_0a_2^3\zeta\xi {v}_2I_6, 
\label{fyfipo}
\end{equation}

\begin{equation} -\frac{\partial {\Psi}_{p_0}}{\partial {v}_3}=
2\pi p_0a_2^3\zeta\xi {v}_3I_7. 
\label{fzfipo}
\end{equation}

\noindent Proceeding in a similar way with the term 
${\Psi}_{p_1}$({\boldmath $v$}) in Eq.~(\ref{pott}), we obtain

\begin{equation} -\frac{\partial {\Psi}_{p_1}}{\partial {v}_1}=
2\pi p_1a_2^4\zeta\xi {v}_1\int_{u(m_1)}^{\lambda} \frac{m^2(u){\rm d}u}
{(a_2^2+u)\sqrt{Au^2+Bu+C}}, 
\label{fxfip1}
\end{equation}

\begin{equation} -\frac{\partial {\Psi}_{p_1}}{\partial {v}_2}=
2\pi p_1a_2^4\zeta\xi {v}_2\int_{u(m_1)}^{\lambda} \frac{m^2(u){\rm d}u}
{(\zeta^2a_2^2+u)\sqrt{Au^2+Bu+C}},
\label{fyfip1}
\end{equation}

\begin{equation} -\frac{\partial {\Psi}_{p_1}}{\partial {v}_3}=
2\pi p_1a_2^4\zeta\xi {v}_3\int_{u(m_1)}^{\lambda} \frac{m^2(u){\rm d}u}
{(\xi^2a_2^2+u)\sqrt{Au^2+Bu+C}}.
\label{fzfip1}
\end{equation}

\noindent The three integrals resulting in each of the
Eqs. (\ref{fxfip1}), (\ref{fyfip1}), (\ref{fzfip1}), are also
computed with standard tables. The total components of the force field
due to the shell are given by adding Eq. (\ref{fxfipo}) with 
Eq. (\ref{fxfip1}), Eq. (\ref{fyfipo}) with Eq. (\ref{fyfip1}),
and Eq. (\ref{fzfipo}) with Eq. (\ref{fzfip1}).\\

\noindent (b) {\boldmath $v$} inside the shell, i.e.
$a$({\boldmath $v$}) $\epsilon [a_1,a_2]$.\\

\noindent In this case all the expressions obtained in part (a) are
 directly
employed, but now instead of the limits of integration $\lambda$
these limits are changed to zero.
If $a$({\boldmath $v$}) = $a_1$, all the force
components are zero.\\

\noindent (c) {\boldmath $v$} inside the cavity of the shell, i.e.
$a$({\boldmath $v$})$< a_1$.
The force components are zero.

\section{Force field of an oblate shell in velocity space with linear
similar density}
\label{ap2}

\noindent (a) {\boldmath $v$} external to the shell, i.e.
$a$({\boldmath $v$})$\geq a_2$.\\

\noindent In the oblate system we take $a=b$, i.e. $\zeta=1$, and
define $R=\sqrt{{v}_1^2+{v}_2^2}$, the eccentricity of the
oblate shell $e_{\rm o}=\sqrt{1-\xi^2}$, and find the solutions
$\beta_1$, $\beta_2$ of the equations

\begin{equation} \begin{array}{rcl}
R^2\sin^2\beta_1+{v}_3^2\tan^2\beta_1 &=& a_1^2e_{\rm o}^2, \\
R^2\sin^2\beta_2+{v}_3^2\tan^2\beta_2 &=& a_2^2e_{\rm o}^2. \end{array}
\label{ecb1b2} \end{equation}

\noindent As in Appendix~\ref{ap1}, the potential is expressed in the form
$\Psi$({\boldmath $v$})= ${\Psi}_{p_0}$({\boldmath $v$})+
${\Psi}_{p_1}$({\boldmath $v$}). The Cartesian components of force
corresponding to ${\Psi}_{p_0}$({\boldmath $v$}) are

\begin{equation} -\frac{\partial {\Psi}_{p_0}}{\partial {v}_1}=
\frac{2\pi p_0\xi {v}_1}{e_{\rm o}^3}\left[F_1(\beta_2)-F_1(\beta_1) \right],
\label{fxfipoob}
\end{equation}

\begin{equation} -\frac{\partial {\Psi}_{p_0}}{\partial {v}_2}=
\frac{2\pi p_0\xi {v}_2}{e_{\rm o}^3}\left[F_1(\beta_2)-F_1(\beta_1) \right], 
\label{fyfipoob}
\end{equation}

\begin{equation} -\frac{\partial {\Psi}_{p_0}}{\partial {v}_3}=
\frac{2\pi p_0\xi {v}_3}{e_{\rm o}^3}\left[F_2(\beta_2)-F_2(\beta_1) \right], 
\label{fzfipoob}
\end{equation}

\noindent with the functions 
$F_1(\beta)$=$\sin\beta\cos\beta-\beta$,
$F_2(\beta)$=$2(\beta-\tan\beta)$.

\noindent The corresponding Cartesian components of force for
${\Psi}_{p_1}$({\boldmath $v$}) are

\begin{equation} -\frac{\partial {\Psi}_{p_1}}{\partial {v}_1}=
2\pi p_1\xi {v}_1\left[F_3(\beta_2)-F_3(\beta_1) \right],
\label{fxfip1ob}
\end{equation}

\begin{equation} -\frac{\partial {\Psi}_{p_1}}{\partial {v}_2}=
2\pi p_1\xi {v}_2\left[F_3(\beta_2)-F_3(\beta_1) \right],
\label{fyfip1ob}
\end{equation}

\begin{equation} -\frac{\partial {\Psi}_{p_1}}{\partial {v}_3}=
2\pi p_1\xi \left[F_4(\beta_2)-F_4(\beta_1) \right],
\label{fzfip1ob}
\end{equation}

\noindent there is no external factor ${v}_3$ in Eq.~(\ref{fzfip1ob})
and the functions $F_3, F_4$ are given by

\begin{equation} \begin{array}{rcl}
F_3(\beta) & = & \frac{1}{e_{\rm o}^4}(F_5(\beta)-F_6(\beta)), \\
F_4(\beta) & = & \frac{1}{e_{\rm o}^4}(-zF_5(\beta)+F_7(\beta)), 
\end{array}
\label{F34}
\end{equation}

\noindent with the functions $F_5, F_6, F_7$ being 

\begin{equation} F_5(\beta)=2\sqrt{R^2\cos^2\beta+{v}_3^2}+
{v}_3\ln\frac{\sqrt{R^2\cos^2\beta+{v}_3^2}-{v}_3}{\sqrt{R^2\cos^2\beta+{v}_3^2}+{v}_3},
\label{F5}
\end{equation}

\begin{equation} F_6(\beta)=\frac{2(R^2\cos^2\beta+{v}_3^2)^{3/2}}{3R^2}, 
\label{F6}
\end{equation}

\begin{equation} F_7(\beta)=-\frac{{v}_3\sqrt{R^2\cos^2\beta+{v}_3^2}}{
\cos^2\beta}+\frac{1}{2}R^2\ln\frac{\sqrt{R^2\cos^2\beta+{v}_3^2}-{v}_3}
{\sqrt{R^2\cos^2\beta+{v}_3^2}+{v}_3}.
\label{F7}
\end{equation}

\noindent The total components of the force field due to the shell are
given by adding Eq. (\ref{fxfipoob}) with
Eq. (\ref{fxfip1ob}), Eq. (\ref{fyfipoob}) with Eq. (\ref{fyfip1ob}),
and Eq. (\ref{fzfipoob}) with Eq. (\ref{fzfip1ob}).\\

\noindent (b) {\boldmath $v$} inside the shell, i.e.
$a$({\boldmath $v$}) $\epsilon [a_1,a_2]$.\\

\noindent In this situation, $\beta_2$ is given by
$\beta_2=\arcsin e_{\rm o}$. If $a$({\boldmath $v$})$> a_1$ and
$a_1 \neq 0$, $\beta_1$ is again the solution of the first equation in
(\ref{ecb1b2}), with $\beta_1=0$ in the case $a_1=0$. All the
expressions in part (a) are employed to compute the
force field. If $a$({\boldmath $v$}) = $a_1$, for any value of
$a_1$ all the force components are zero.\\

\noindent (c) {\boldmath $v$} inside the cavity of the shell, i.e.
$a$({\boldmath $v$})$< a_1$.
The force components are zero.

\section{Force field of a prolate shell in velocity space with linear
similar density}
\label{ap3}

\noindent (a) {\boldmath $v$} external to the shell, i.e.
$a$({\boldmath $v$})$\geq a_2$.\\

\noindent In this case we take $b=c$, and define
$\mathcal{R}=\sqrt{{v}_2^2+{v}_3^2}$, the eccentricity of the
prolate shell $e_{\rm p}=\sqrt{1-\xi^2}$, and find the solutions
$\beta_1$, $\beta_2$ of the equations

\begin{equation} \begin{array}{rcl}
{v}_1^2\sin^2\beta_1+\mathcal{R}^2\tan^2\beta_1 &=& a_1^2e_{\rm p}^2, \\
{v}_1^2\sin^2\beta_2+\mathcal{R}^2\tan^2\beta_2 &=& a_2^2e_{\rm p}^2. \end{array}
\label{eccb1b2} \end{equation}

\noindent As in Appendix~\ref{ap1}, the potential is expressed in the form
$\Psi$({\boldmath $v$})= ${\Psi}_{p_0}$({\boldmath $v$})+
${\Psi}_{p_1}$({\boldmath $v$}). The Cartesian components of force
corresponding to ${\Psi}_{p_0}$({\boldmath $v$}) are

\begin{equation} -\frac{\partial {\Psi}_{p_0}}{\partial {v}_1}=
\frac{2\pi p_0\xi^2 {v}_1}{e_{\rm p}^3}\left[H_1(\beta_2)-H_1(\beta_1) \right],
\label{fxfipopro}
\end{equation}

\begin{equation} -\frac{\partial {\Psi}_{p_0}}{\partial {v}_2}=
\frac{2\pi p_0\xi^2 {v}_2}{e_{\rm p}^3}\left[H_2(\beta_2)-H_2(\beta_1) \right],
\label{fyfipopro}
\end{equation}

\begin{equation} -\frac{\partial {\Psi}_{p_0}}{\partial {v}_3}=
\frac{2\pi p_0\xi^2 {v}_3}{e_{\rm p}^3}\left[H_2(\beta_2)-H_2(\beta_1) \right],
\label{fzfipopro}
\end{equation}

\noindent with the functions 
$H_1(\beta)$=2$(\sin\beta -\ln[(1+\sin\beta)/\cos\beta])$,
$H_2(\beta)$=$\ln[(1+\sin\beta)/\cos\beta]-\sin\beta/\cos^2\beta$.

\noindent The corresponding Cartesian components of force for
${\Psi}_{p_1}$({\boldmath $v$}) are

\begin{equation} -\frac{\partial {\Psi}_{p_1}}{\partial {v}_1}=
\frac{2\pi p_1\xi^2H_3}{e_{\rm p}^4},
\label{fxfip1pro}
\end{equation}

\begin{equation} -\frac{\partial {\Psi}_{p_1}}{\partial {v}_2}=
\frac{2\pi p_1\xi^2{v}_2H_4}{e_{\rm p}^4},
\label{fyfip1pro}
\end{equation}

\begin{equation} -\frac{\partial {\Psi}_{p_1}}{\partial {v}_3}=
\frac{2\pi p_1\xi^2{v}_3H_4}{e_{\rm p}^4},
\label{fzfip1pro}
\end{equation}

\noindent there is no factor ${v}_1$ in Eq.~(\ref{fxfip1pro}) and
the functions $H_3, H_4$ are given by

\begin{eqnarray} H_3 &=& -{v}_1 \left [(2+\cos^2\beta_2)\frac{S_2^{1/2}}{\cos\beta_2}-(2+\cos^2\beta_1)\frac{S_1^{1/2}}{\cos\beta_1}\right]\pm
\nonumber \\
& & \pm (\mathcal{R}^2-2{v}_1^2)L,
\label{H3}
\end{eqnarray}

\noindent with + sign if ${v}_1>0$ and $-$ sign if ${v}_1<0$,

\begin{eqnarray} H_4 &=& -\frac{2}{3\mathcal{R}^2} \left [\frac{S_2^{3/2}}{\cos^3\beta_2}-\frac{S_1^{3/2}}{\cos^3\beta_1}\right]+
2\left [\frac{S_2^{1/2}}{\cos\beta_2}-\frac{S_1^{1/2}}{\cos\beta_1}\right]+
\nonumber \\
& & + 2|{v}_1|L, 
\label{H4}
\end{eqnarray}

\noindent with the functions

\begin{equation} \begin{array}{rcl}
S_1 &=& {v}_1^2\cos^2\beta_1+\mathcal{R}^2, \\
S_2 &=& {v}_1^2\cos^2\beta_2+\mathcal{R}^2,\\
L &=& \ln \frac{S_2^{1/2}-|{v}_1|\cos\beta_2}{S_1^{1/2}-|{v}_1|\cos\beta_1}.
\end{array}
\label{S12L} \end{equation}

\noindent The total components of the force field due to the shell are 
given by adding Eq. (\ref{fxfipopro}) with
Eq. (\ref{fxfip1pro}), Eq. (\ref{fyfipopro}) with Eq. (\ref{fyfip1pro}),
and Eq. (\ref{fzfipopro}) with Eq. (\ref{fzfip1pro}).\\

\noindent (b) {\boldmath $v$} inside the shell, i.e.
$a$({\boldmath $v$}) $\epsilon [a_1,a_2]$.\\

\noindent In this situation, $\beta_2$ is given by
$\beta_2=\arcsin e_{\rm p}$. If $a$({\boldmath $v$})$> a_1$ and
$a_1 \neq 0$, $\beta_1$ is again the solution of the first equation in
(\ref{eccb1b2}), with $\beta_1=0$ in the case $a_1=0$. All the
expressions in part (a) are employed to compute the
force field. If $a$({\boldmath $v$}) = $a_1$, for any value of
$a_1$ all the force components are zero.\\

\noindent (c) {\boldmath $v$} inside the cavity of the shell, i.e.
$a$({\boldmath $v$})$< a_1$.
The force components are zero.

\section{Comparison of perigalactic and apogalactic distances in two
Galactic models, without the dynamical friction effect}
\label{ap4}

We compare here the mean perigalactic and apogalactic distances
reached by some globular clusters, computed without the dynamical
friction effect, as obtained with the Galactic model employed in this
work, and with Galactic Model 2 used by \citet{2018A&A...616A..12G} and
in the Holger Baumgardt compilation cited in Section~\ref{orbitas},
which is the corresponding Model I of \citet{2013A&A...549A.137I}.
These two axisymmetric Galactic models which are employed in this
comparison are slight modifications of the original model presented by
\citet{1991RMxAA..22..255A}. Thus, under equal orbital initial
conditions, no great differences are expected in the computations
of Galactic orbits using these models. Table~\ref{cumulos.ax} shows
the mean values of $<\!r_{\rm min}\!>$ and $<\!r_{\rm max}\!>$
in some clusters analysed in the present work, obtained with our
model and with Irrgang's model. 
Although both Galactic models and the orbital initial conditions
(dependent on different assumed positions of the Sun and its peculiar
velocity, motion of the local standard of rest, distances and radial
velocities from the Sun, etc.) are not exactly the same in this work
and in the analyses presented by
\citet{2018A&A...616A..12G} and Holger Baumgardt compilation,
the corresponding listed values of the
perigalactic and apogalactic distances compare well.

\begin{table}
 \caption{Comparison in some clusters. Columns with label 1 give
 properties obtained with the Galactic model employed in this work,
 those with label 2 and 3 list respectively results from
 \citet{2018A&A...616A..12G} and the Holger Baumgardt compilation,
 using the Galactic Model I of \citet{2013A&A...549A.137I}.} 
 \label{cumulos.ax}
 \begin{tabular}{cccccccc}
  \hline
 & \multicolumn{3}{c}{$<\!r_{\rm min}\!>$ (kpc)} & & \multicolumn{3}{c}{$<\!r_{\rm max}\!>$ (kpc)} \\
         &   1   &   2   & 3 & &  1  &   2   & 3 \\ 
  \hline
 NGC 104  & 5.76  & 5.68  & 5.47  & & 7.45  & 7.68  & 7.51    \\ 
 NGC 4372 & 2.96  & 3.14  & 2.92  & & 7.25  & 7.40  & 7.24    \\
 NGC 4833 & 0.97  & 0.93  & 0.85  & & 7.24  & 7.86  & 7.65    \\
 NGC 5139 & 1.87  & 1.29  & 1.50  & & 6.88  & 7.26  & 6.94    \\
 NGC 5927 & 4.24  & 4.31  & 4.03  & & 5.67  & 5.19  & 5.46    \\
 NGC 5946 & 0.61  & 0.76  & 0.49  & & 4.97  & 5.72  & 5.18    \\
 NGC 5986 & 0.62  & 0.52  & 0.56  & & 5.24  & 4.95  & 5.05    \\
 NGC 6093 & 0.84  & 0.60  & 0.87  & & 4.39  & 3.89  & 4.07    \\
 NGC 6121 & 0.90  & 0.41  & 0.68  & & 5.97  & 6.08  & 6.44    \\
 NGC 6144 & 1.86  & 2.39  & 1.90  & & 3.24  & 2.96  & 3.13    \\
 NGC 6171 & 0.89  & 0.94  & 1.22  & & 4.38  & 3.74  & 3.83    \\
 NGC 6218 & 2.08  & 2.41  & 2.15  & & 4.80  & 5.00  & 4.71    \\
 NGC 6235 & 3.46  & 3.15  & 3.49  & & 8.57  & 5.04  & 7.47    \\
 NGC 6254 & 1.93  & 2.30  & 1.92  & & 4.56  & 5.28  & 4.59    \\
 NGC 6266 & 0.85  & 0.69  & 0.86  & & 2.33  & 2.36  & 2.42    \\
 NGC 6273 & 1.09  & 1.32  & 1.21  & & 3.32  & 3.38  & 3.28    \\
 NGC 6284 & 0.89  & 1.07  & 0.99  & & 6.56  & 7.31  & 6.43    \\
 NGC 6287 & 0.56  & 1.24  & 0.88  & & 4.65  & 4.74  & 4.16    \\
 NGC 6293 & 0.47  & 0.47  & 0.47  & & 2.74  & 2.81  & 2.74    \\
 NGC 6304 & 1.45  & 1.87  & 1.55  & & 2.66  & 3.13  & 2.69    \\
 NGC 6316 & 1.09  & 0.42  & 0.90  & & 3.89  & 2.59  & 3.92    \\
 NGC 6325 & 0.95  & 1.15  & 1.04  & & 1.29  & 1.35  & 1.31    \\
 NGC 6333 & 0.91  & 1.45  & 1.22  & & 7.07  & 3.52  & 6.16    \\
 NGC 6342 & 0.74  & 0.97  & 0.83  & & 1.67  & 1.66  & 1.67    \\
 NGC 6352 & 3.11  & 3.20  & 3.08  & & 4.11  & 4.36  & 3.93    \\
 NGC 6356 & 3.72  & 2.64  & 3.53  & & 8.94  & 7.55  & 8.97    \\
 NGC 6362 & 2.80  & 5.12  & 2.65  & & 5.22  & 6.83  & 5.24    \\
 NGC 6366 & 2.09  & 3.56  & 2.21  & & 5.87  & 5.38  & 5.71    \\
 NGC 6380 & 0.33  & 0.39  & 0.30  & & 2.02  & 2.90  & 2.10    \\
 NGC 6388 & 1.28  & 0.74  & 1.15  & & 3.83  & 3.18  & 4.07    \\
 NGC 6397 & 2.41  & 2.90  & 2.61  & & 6.28  & 6.59  & 6.30    \\
 NGC 6401 & 0.23  & 2.40  & 0.29  & & 1.31  & 3.90  & 1.55    \\
 NGC 6402 & 0.62  & 0.51  & 0.53  & & 4.24  & 4.32  & 4.28    \\
 NGC 6440 & 0.16  & 0.23  & 0.22  & & 1.30  & 1.30  & 1.26    \\
 NGC 6441 & 1.65  & 0.80  & 1.59  & & 4.72  & 3.43  & 4.71    \\
 NGC 6453 & 0.37  & 1.22  & 0.48  & & 2.68  & 3.62  & 2.58    \\
 NGC 6496 & 2.45  & 3.79  & 2.62  & & 5.33  & 8.11  & 4.62    \\
 NGC 6517 & 0.45  & 0.49  & 0.44  & & 3.26  & 4.04  & 3.36    \\
 NGC 6522 & 0.20  & 0.31  & 0.45  & & 1.41  & 1.23  & 1.31    \\
 NGC 6528 & 0.40  & 0.39  & 0.42  & & 0.72  & 1.03  & 0.92    \\
 NGC 6535 & 1.18  & 1.10  & 0.97  & & 4.31  & 4.52  & 4.52    \\
 NGC 6539 & 2.02  & 1.95  & 2.11  & & 3.37  & 3.29  & 3.41    \\
 NGC 6541 & 1.59  & 1.44  & 1.56  & & 3.62  & 3.70  & 3.59    \\
 NGC 6544 & 0.66  & 0.39  & 0.60  & & 5.16  & 5.23  & 5.54    \\
 NGC 6626 & 0.47  & 0.58  & 0.59  & & 3.05  & 3.15  & 3.03    \\
 NGC 6637 & 0.36  & 0.32  & 0.39  & & 2.06  & 1.93  & 2.06    \\
 NGC 6656 & 3.04  & 3.21  & 2.96  & & 9.82  & 9.92  & 9.59    \\
 NGC 6681 & 0.54  & 1.16  & 0.86  & & 5.49  & 4.31  & 4.97    \\
 NGC 6752 & 3.49  & 3.64  & 3.37  & & 5.62  & 5.72  & 5.48    \\
 NGC 6809 & 1.45  & 1.59  & 1.57  & & 5.78  & 5.82  & 5.62    \\
 NGC 6838 & 4.86  & 4.99  & 4.81  & & 7.08  & 7.30  & 7.12    \\
  \hline
 \end{tabular}
\end{table}

\end{document}